\documentclass[12pt]{article}
\pdfoutput=1 

\usepackage[utf8]{inputenc}

\usepackage{amsmath,amsfonts,amssymb,epsfig}
\usepackage{slashed}
\usepackage{cite}
\usepackage{multirow}
\usepackage{cite}

\usepackage{hyperref}

\usepackage{calc}
\usepackage{rotating}
\usepackage[english]{babel}

\usepackage{graphicx}
\usepackage{subfig}
\usepackage{float}
\usepackage{amsmath}
\usepackage{amssymb}
\usepackage{amsthm}
\usepackage{latexsym}
\usepackage{dcolumn}
\usepackage{geometry}
\usepackage{color}
\usepackage{hyperref}
\usepackage{mciteplus}
\usepackage{fancyhdr}

\usepackage[normalem]{ulem}
\usepackage[utf8]{inputenc}

\hypersetup{
    colorlinks,
    linkcolor={black},
    citecolor={black},
    urlcolor={black}
}

\def\gtwid{\mathrel{\raise.3ex\hbox{$>$\kern-.75em\lower1ex\hbox{$\sim
$}}}}
\def\vio{\mathrel{\hbox{$E$\kern-.60em\hbox{$/
$}}}}
\newcommand{\newc}{\newcommand*}

\long\def\begincomment#1\endcomment{%
        \begingroup\sf\baselineskip12pt#1\endgroup}

\newc{\etal}{\textrm{et al.}} 
\newc{\eg}{\textrm{e.g.}} 
\newc{\ie}{\textrm{i.e.}}
\newc{\etc}{\textrm{etc}}
\newc\vs{\textrm{vs.}}
\newc{\cl}{\rm {C.L.}}

\newc{\ev}{\ensuremath{\,\mathrm{eV}}}
\newc{\kev}{\ensuremath{\,\mathrm{keV}}}
\newc{\mev}{\ensuremath{\,\mathrm{MeV}}}
\newc{\gev}{\ensuremath{\,\mathrm{GeV}}}
\newc{\tev}{\ensuremath{\,\mathrm{TeV}}}
\newc{\MeV}{\mev} 
\newc{\TeV}{\tev}
\newc{\invpb}{\ensuremath{/\text{pb}}}
\newc{\invfb}{\ensuremath{\,\text{fb}^{-1}}}
\newc\nb{\ensuremath{\,\mathrm{nb}}} \newc\pb{\ensuremath{\,\mathrm{pb}}} \newc\fb{\ensuremath{\,\mathrm{fb}}}
\newc\pc{\ensuremath{\,\mathrm{pc}}}
\newc\kpc{\ensuremath{\,\mathrm{kpc}}}
\newc\mpc{\ensuremath{\,\mathrm{Mpc}}}
\newc\ps{\ensuremath{\,\mathrm{ps}}} 
\newc\cmeter{\ensuremath{\,\mathrm{cm}}} 
\newc\meter{\ensuremath{\,\mathrm{m}}} 
\newc\kmeter{\ensuremath{\,\mathrm{km}}}
\newc\second{\ensuremath{\,\mathrm{s}}}
\newc\msecond{\ensuremath{\,\mathrm{ms}}}
\newc\nsecond{\ensuremath{\,\mathrm{ns}}}
\newc\psecond{\ensuremath{\,\mathrm{ps}}}

\newc{\chisqmin}{\ensuremath{\chi^2_{\mathrm{min}}}}
\newc{\Delchisq}{\ensuremath{\Delta\chi^2}}
\newc{\chisq}{\ensuremath{\chi^2}}
\newc{\like}{\ensuremath{\mathcal{L}}}

\newc\lsim{\ensuremath{\mathrel{\rlap{\lower4pt\hbox{\hskip1pt$\sim$}}\raise1pt\hbox{$<$}}}}
\newc\gsim{\ensuremath{\mathrel{\rlap{\lower4pt\hbox{\hskip1pt$\sim$}}\raise1pt\hbox{$>$}}}}
\newc{\VEV}[1]{\ensuremath{\langle #1 \rangle}}
\newc{\dl}{\ensuremath{\stackrel{\leftarrow}{D}}}
\newc{\dr}{\ensuremath{\stackrel{\rightarrow}{D}}}

\newc{\scr}[1]{\ensuremath{\mathcal{#1}}}

\newc{\bcenter}{\begin{center}}    \newc{\ecenter}{\end{center}}
\newc{\bfl}{\begin{flushleft}}    \newc{\efl}{\end{flushleft}}
\newc{\bfr}{\begin{flushright}}    \newc{\efr}{\end{flushright}}

\newc{\bi}{\begin{itemize}}
\newc{\ei}{\end{itemize}}
\newc{\bed}{\begin{description}}
\newc{\eed}{\end{description}}
\newc{\ben}{\begin{enumerate}}
\newc{\een}{\end{enumerate}}

\newc{\be}{\begin{equation}}
\newc{\ee}{\end{equation}}
\newc{\bea}{\begin{eqnarray}}
\newc{\eea}{\end{eqnarray}}
\newc{\ra}{\rightarrow}

\newc{\alphas}{\ensuremath{\alpha_s}}
\newc{\alphatwo}{\ensuremath{\alpha_2}}
\newc{\alphaone}{\ensuremath{\alpha_1}}
\newc{\alphai}[1]{\ensuremath{\alpha_{#1}}}
\newc{\alphaem}{\ensuremath{\alpha_{\mathrm{em}}}}
\newc{\alphaeff}{\ensuremath{\alpha_{\mathrm{eff}}}}
\newc{\sineff}{\ensuremath{\sin \theta_{\mathrm{eff}}}}
\newc{\sinsqeff}{\ensuremath{\sin^2 \theta_{\mathrm{eff}}}}
\newc{\dalphahad}{\ensuremath{\Delta \alpha_{\mathrm{had}}}}
\newc{\yt}{\ensuremath{h_t}} \newc{\yb}{\ensuremath{h_b}} \newc{\ytau}{\ensuremath{h_{\tau}}}
\newc\mz{\ensuremath{M_Z}} 
\newc\mw{\ensuremath{m_W}}
\newc\mZ{\mz}        \newc\mW{\mw}
\newc\mhsm{\ensuremath{ m_{H_{\mathrm{SM}}}}}
\newc{\mtop}{\ensuremath{ m_t}}               \newc{\mtpole}{\ensuremath{ M_t}}
\newc{\mbottom}{\ensuremath{ m_b}} 
\newc{\mtau}{\ensuremath{ m_{\tau}}}
\newc{\mt}{\mtpole}
\newc{\mb}{\mbottom} 
\newc{\rgg}{\ensuremath{R_{h}(\gamma\gamma)}}
\newc{\rzz}{\ensuremath{R_{h}(ZZ)}}
\newc{\rtwogg}{\ensuremath{R_{h_2}(\gamma\gamma)}}
\newc{\rtwozz}{\ensuremath{R_{h_2}(ZZ)}}
\newc{\ronegg}{\ensuremath{R_{h_1}(\gamma\gamma)}}
\newc{\ronezz}{\ensuremath{R_{h_1}(ZZ)}}
\newc{\rsiggg}{\ensuremath{R_{h_\textrm{sig}}(\gamma\gamma)}}
\newc{\rsigzz}{\ensuremath{R_{h_\textrm{sig}}(ZZ)}}
\newc{\llbar}{\ensuremath{\ell\bar{\ell}}}
\newc{\tauptaum}{\ensuremath{ \tau^+\tau^-}}
\newc{\qqbar}{\ensuremath{ q\bar{q}}} \newc{\ppbar}{\ensuremath{ p\bar{p}}}
\newc{\bbbar}{\ensuremath{ b\bar{b}}} \newc{\ttbar}{\ensuremath{ t\bar{t}}}
\newc{\ffbar}{\ensuremath{ f\bar{f}}} \newc{\tautaubar}{\ensuremath{ \tau\bar{\tau}}}
\newc{\mchi}{\ensuremath{m_{\chi}}}
\newc{\squark}{\ensuremath{\tilde{q}}}
\newc{\slepton}{\ensuremath{\tilde{l}}}
\newc{\gluino}{\ensuremath{\tilde{g}}} 
\newc{\mgluino}{\ensuremath{{m_{\gluino}}}}
\newc{\tone}{\ensuremath{{\tilde{t}_1}}}

\newc{\sthw}{\ensuremath{ \sin\theta_W}}              \newc{\cthw}{\ensuremath{\cos\theta_W}}
\newc{\tanthw}{\ensuremath{ \tan\theta_W}}              \newc{\cotthw}{\ensuremath{\cot\theta_W}}
\newc{\ssqthw}{\ensuremath{\sin^2 \theta_W}}
\newc{\msbar}{\ensuremath{\overline{MS}}} \newc{\drbar}{\ensuremath{\overline{DR}}}
\newc{\mtmtsmmsbar}{\ensuremath{ m_t(m_t)^{\msbar}_{{\mathrm{SM}}}}}
\newc{\mtmtsmdrbar}{\ensuremath{ m_t(m_t)^{\drbar}_{{\mathrm{SM}}}}}
\newc{\mtmtmssmdrbar}{\ensuremath{ m_t(m_t)^{\drbar}_{{\mathrm{SUSY}}}}}
\newc{\mbmbmsbar}{\ensuremath{ m_b(m_b)^{\msbar} }}
\newc{\mbmbsmmsbar}{\ensuremath{ m_b(m_b)^{\msbar}_{{\mathrm{SM}}}}}
\newc{\mbmzsmmsbar}{\ensuremath{ m_b(\mz)^{\msbar}_{{\mathrm{SM}}}}}
\newc{\mbmzsmdrbar}{\ensuremath{ m_b(\mz)^{\drbar}_{{\mathrm{SM}}}}}
\newc{\mbmzmssmdrbar}{\ensuremath{ m_b(\mz)^{\drbar}_{{\mathrm{SUSY}}}}}
\newc{\mtaumzsmmsbar}{\ensuremath{ m_{\tau}(\mz)^{\msbar}_{{\mathrm{SM}}}}}
\newc{\mtaumzsmdrbar}{\ensuremath{ m_{\tau}(\mz)^{\drbar}_{{\mathrm{SM}}}}}
\newc{\mtaumzmssmdrbar}{\ensuremath{ m_{\tau}(\mz)^{\drbar}_{{\mathrm{SUSY}}}}}
\newc{\alphasmzms}{\ensuremath{\alpha_s(M_Z)^{\overline{MS}}}}
\newc{\alphaimzms}[1]{\ensuremath{\alpha_{#1}(M_Z)^{\overline{MS}}}}
\newc{\alphaemmz}{\ensuremath{\alpha_{\mathrm{em}}(M_Z)^{\overline{MS}}}}

\newc{\mzero}{\ensuremath{{m_0}}}
\newc{\mhalf}{\ensuremath{ m_{1/2}}}
\newc{\tanb}{\ensuremath{\tan\beta}}
\newc{\azero}{\ensuremath{ A_0}}
\newc{\bzero}{\ensuremath{ B_0}}
\newc{\signmu}{\ensuremath{\rm{sgn}\,\mu}}
\newc{\mueff}{\ensuremath{\mu_{\rm{eff}}}}
\newc{\lam}{\ensuremath{{\lambda}}}
\newc{\kap}{\ensuremath{{\kappa}}}
\newc{\alam}{\ensuremath{{A_{\lambda}}}}
\newc{\akap}{\ensuremath{{A_{\kappa}}}}
\newc{\hs}{\ensuremath{ H_s}}      
\newc{\mhs}{\ensuremath{ m_{H_s}}} 
\newc{\mgut}{\ensuremath{ M_{\rm GUT}}}
\newc{\mplanck}{\ensuremath{ M_{\rm P}}}      \newc{\mpl}{\ensuremath{ M_{\rm Pl}}}
\newc{\msusy}{\ensuremath{ M_{\rm SUSY}}}      \newc{\ms}{\ensuremath{ M_{\rm S}}}
 \newc{\mhl}{\ensuremath{m_\hl}} 
 \newc{\mhone}{\ensuremath{m_{h_1}}} 
 \newc{\mhtwo}{\ensuremath{m_{h_2}}} 
 \newc{\mglu}{\ensuremath{m_{\tilde g}}} 
 \newc{\mul}{\ensuremath{m_{\tilde{u}_L}}} 
 \newc{\mtone}{\ensuremath{m_{\tilde{t}_1}}} 
 \newc{\ma}{\ensuremath{m_A}} 
 \newc{\maone}{\ensuremath{m_{a_1}}} 
 \newc{\matwo}{\ensuremath{m_{a_2}}}
 \newc{\hone}{\ensuremath{h_1}}
 \newc{\htwo}{\ensuremath{h_2}}
 \newc{\aone}{\ensuremath{a_1}}
 \newc{\atwo}{\ensuremath{a_2}}
 \newc{\mhu}{\ensuremath{ m_{H_u}}}       
 \newc{\mhd}{\ensuremath{ m_{H_d}}}
 \newc{\mhusq}{\ensuremath{ m_{H_u}^2}}       
 \newc{\mhdsq}{\ensuremath{ m_{H_d}^2}}
 \newc{\mhuew}{\ensuremath{ m^{\ast}_{H_u}}}       
 \newc{\mhdew}{\ensuremath{ m^{\ast}_{H_d}}}
 \newc{\mhuewsq}{\ensuremath{ m^{\ast\, 2}_{H_u}}}       
 \newc{\mhdewsq}{\ensuremath{ m^{\ast\, 2}_{H_d}}}
 \newc{\hu}{\ensuremath{ H_u}}       
 \newc{\hd}{\ensuremath{ H_d}}
 \newc{\barmhu}{\ensuremath{ \bar{m}_{H_u}}}
 \newc{\barmhd}{\ensuremath{ \bar{m}_{H_d}}}

 \newc{\mqthree}{\ensuremath{m_{\widetilde{Q}_3}^2}}
 \newc{\muthree}{\ensuremath{m_{\tilde{u}_3}^2}}
 \newc{\mdthree}{\ensuremath{m_{\tilde{d}_3}^2}}
 \newc{\mlthree}{\ensuremath{m_{\widetilde{L}_3}^2}}
 \newc{\methree}{\ensuremath{m_{\tilde{e}_3}^2}}
 \newc{\mqtwo}{\ensuremath{m_{\widetilde{Q}_2}^2}}
 \newc{\mutwo}{\ensuremath{m_{\tilde{u}_2}^2}}
 \newc{\mdtwo}{\ensuremath{m_{\tilde{d}_2}^2}}
 \newc{\mltwo}{\ensuremath{m_{\widetilde{L}_2}^2}}
 \newc{\metwo}{\ensuremath{m_{\tilde{e}_2}^2}}
 \newc{\mqone}{\ensuremath{m_{\widetilde{Q}_1}^2}}
 \newc{\muone}{\ensuremath{m_{\tilde{u}_1}^2}}
 \newc{\mdone}{\ensuremath{m_{\tilde{d}_1}^2}}
 \newc{\mlone}{\ensuremath{m_{\widetilde{L}_1}^2}}
 \newc{\meone}{\ensuremath{m_{\tilde{e}_1}^2}}
 \newc{\msmul}{\ensuremath{m_{\tilde{\mu}_L}}}
 \newc{\msmur}{\ensuremath{m_{\tilde{\mu}_R}}}
 \newc{\msneumu}{\ensuremath{m_{\tilde{\nu}_{\mu}}}}
 \newc{\mone}{\ensuremath{M_1}}
 \newc{\monesq}{\ensuremath{M_1^2}}
 \newc{\mtwo}{\ensuremath{M_2}}
 \newc{\mtwosq}{\ensuremath{M_2^2}}
 \newc{\mthree}{\ensuremath{M_3}}
 \newc{\mthreesq}{\ensuremath{M_3^2}}
 \newc{\atau}{\ensuremath{{A_{\tau}}}}
 \newc{\at}{\ensuremath{{A_{t}}}}
 \newc{\ab}{\ensuremath{{A_{b}}}}
 \newc{\atausq}{\ensuremath{{A_{\tau}^2}}}
 \newc{\atsq}{\ensuremath{{A_{t}^2}}}
 \newc{\absq}{\ensuremath{{A_{b}^2}}}

 \newc{\dmzero}{\ensuremath{\Delta{_{m_0}}}}
 \newc{\dmhalf}{\ensuremath{\Delta{_{m_{1/2}}}}}
 \newc{\dmu}{\ensuremath{\Delta{_{\mu}}}}

 \newc{\pten}{\ensuremath{\psi_{10}}}
 \newc{\ffive}{\ensuremath{\phi_{5}}}
 \newc{\hfive}{\ensuremath{h_{5}}}
 \newc{\hbfive}{\ensuremath{h_{\bar{5}}}}
 \newc{\thet}{\ensuremath{\theta_{50}}}
 \newc{\thetb}{\ensuremath{\theta_{\,\overline{50}}}}
 \newc{\ptenhat}{\ensuremath{\hat{\psi}_{10}}}
 \newc{\ffivehat}{\ensuremath{\hat{\phi}_{5}}}
 \newc{\hfivehat}{\ensuremath{\hat{h}_{5}}}
 \newc{\hbfivehat}{\ensuremath{\hat{h}_{\bar{5}}}}
 \newc{\thethat}{\ensuremath{\hat{\theta}_{50}}}
 \newc{\thetbhat}{\ensuremath{\hat{\theta}_{\,\overline{50}}}}
 \newc{\si}{\ensuremath{\Sigma}}
 \newc{\mfive}{\ensuremath{m_5^2}}
 \newc{\mten}{\ensuremath{m_{10}^2}}
 \newc{\dfive}{\ensuremath{\Delta^2_5}}
 \newc{\dbfive}{\ensuremath{\Delta^2_{\bar{5}}}}
 \newc{\dfifty}{\ensuremath{\Delta^2_{50}}}
 \newc{\dfiftyb}{\ensuremath{\Delta^2_{\,\overline{50}}}}
 \newc{\msi}{\ensuremath{m_{\Sigma}^2}}
 \newc{\lamh}{\ensuremath{\lambda_{H}}}
 \newc{\lamhb}{\ensuremath{\lambda_{\bar{H}}}}
 \newc{\ah}{\ensuremath{A_{H}}}
 \newc{\ahb}{\ensuremath{A_{\bar{H}}}}
 \newc{\lams}{\ensuremath{\lambda_{S}}}
 \newc{\as}{\ensuremath{A_{S}}}
 \newc{\lamsig}{\ensuremath{\lambda_{\si}}}
 \newc{\asig}{\ensuremath{A_{\si}}}

 \newc{\msten}{\ensuremath{m_{16}^2}}
 \newc{\mhun}{\ensuremath{m_{126}^2}}
 \newc{\mhunb}{\ensuremath{m_{\bar{126}}^2}}
 \newc{\mthun}{\ensuremath{m_{210}^2}}
 \newc{\ahun}{\ensuremath{A_{\bar{126}}}}
 \newc{\yhun}{\ensuremath{Y_{\bar{126}}}}
 \newc{\aten}{\ensuremath{A_{10}}}
 \newc{\yten}{\ensuremath{Y_{10}}}
 \newc{\alone}{\ensuremath{A_{\lambda_1}}}
 \newc{\altwo}{\ensuremath{A_{\lambda_2}}}
 \newc{\althree}{\ensuremath{A_{\lambda_3}}}
 \newc{\althreeb}{\ensuremath{A_{\bar{\lambda_3}}}}
 \newc{\lone}{\ensuremath{\lambda_1}}
 \newc{\ltwo}{\ensuremath{\lambda_2}}
 \newc{\lthree}{\ensuremath{\lambda_3}}
 \newc{\lthreeb}{\ensuremath{\bar{\lambda_3}}}

\newc{\sigsip}{\ensuremath{\sigma^{\rm SI}_{p}}}	\newc{\sigsin}{\ensuremath{\sigma^{\rm SI}_{n}}}
\newc{\sigsdp}{\ensuremath{\sigma^{\rm SD}_{p}}}	\newc{\sigsdn}{\ensuremath{\sigma^{\rm SD}_{n}}}
\newc{\sigsi}{\ensuremath{\sigma^{\rm SI}}}	\newc{\sigsd}{\ensuremath{\sigma^{\rm SD}}}
\newc{\sigv}{\ensuremath{\sigma v}}
\newc{\abund}{\ensuremath{ \Omega h^2}}
\newc{\omegadm}{\ensuremath{ \Omega_{{\rm DM}}}}     \newc{\abunddm}{\ensuremath{ \Omega_{{\rm DM}} h^2}} 
\newc{\omegam}{\ensuremath{ \Omega_{{\rm m}}}}       \newc{\abundm}{\ensuremath{ \Omega_{{\rm m}} h^2}}
\newc{\omegab}{\ensuremath{ \Omega_{{\rm b}}}}	\newc{\abundb}{\ensuremath{ \Omega_{{\rm b}} h^2}}
\newc{\omegatot}{\ensuremath{ \Omega_{{\rm TOT}}}}
\newc{\omegacdm}{\ensuremath{ \Omega_{{\rm CDM}}}}   \newc{\abundcdm}{\ensuremath{ \Omega_{{\rm CDM}} h^2}}
\newc{\omegalambda}{\ensuremath{ \Omega_{\Lambda}}} \newc{\abundlambda}{\ensuremath{ \Omega_{\Lambda} h^2}}
\newc{\omegarad}{\ensuremath{ \Omega_{{\rm rad}}}}  \newc{\abundrad}{\ensuremath{ \Omega_{{\rm rad}} h^2}}
\newc{\rhocrit}{\ensuremath{ \rho_{\rm crit}}}
\newc{\rhochi}{\ensuremath{ \rho_{\chi}}}
\newc{\abunchi}{\ensuremath{\Omega_\chi h^2}}
\newc{\abundlsp}{\ensuremath{\Omega_{\rm LSP}h^2}}

\newc{\amu}{\ensuremath{ a_{\mu}}}        \newc{\amususy}{\ensuremath{ a_{\mu}^{\mathrm{SUSY}}}}
\newc{\amuexpt}{\ensuremath{ a_{\mu}^{\mathrm{expt}}}}        \newc{\amusm}{\ensuremath{ a_{\mu}^{\mathrm{SM}}}}
\newc\deltaamu{\ensuremath{\Delta a_{\mu}}} \newc{\deltaamususy}{\ensuremath{\delta a_{\mu}^{\mathrm{SUSY}}}}
\newc\gmtwo{\ensuremath{ (g-2)_{\mu}}} 
\newc{\deltagmtwomususy}{\ensuremath{\delta\left(g-2\right)_{\mu}^{\mathrm{SUSY}}}}
\newc{\deltagmtwomu}{\ensuremath{\delta\left(g-2\right)_{\mu}}}
\newc\BR{\ensuremath{\textrm{BR}}}
\newc\bsgamma{\ensuremath{ b\rightarrow s \gamma }}
\newc\bxsgamma{\ensuremath{\overline{B}\rightarrow X_{s}\gamma}}
\newc\brbsgamma{\ensuremath{\BR\left(\bsgamma\right)}}
\newc\brbxsgamma{\ensuremath{\BR\left(\bxsgamma\right)}}
\newc\bsmumu{\ensuremath{B_s\to\mu^+\mu^-}}
\newc\brbsmumu{\ensuremath{\BR\left(B_s\to\mu^+\mu^-\right)}}
\newc\bdmmumu{\ensuremath{\overline{B}_d\to\mu^+\mu^-}}
\newc\bbbarmix{\ensuremath{\overline{B}_s\mbox{-}B_s}}      
\newc\delmbs{\ensuremath{\Delta M_{B_s}}}
\newc{\butaunu}{\ensuremath{B_u \rightarrow \tau \nu}}
\newc{\brbutaunu}{\ensuremath{\BR\left(B_u \rightarrow \tau \nu\right)}}

\newc{\brmuegamma}{\ensuremath{\BR\left(\mu^{\pm}\rightarrow e^{\pm}\gamma\right)}}
\newc{\brtauegamma}{\ensuremath{\BR\left(\tau^{\pm}\rightarrow e^{\pm}\gamma\right)}}
\newc{\brtaumugamma}{\ensuremath{\BR\left(\tau^{\pm}\rightarrow \mu^{\pm}\gamma\right)}}
\newc{\brmuthreee}{\ensuremath{\BR\left(\mu^{\pm}\rightarrow e^{\pm}e^+e^-\right)}}
\newc{\brtauthreee}{\ensuremath{\BR\left(\tau^{\pm}\rightarrow e^{\pm}e^+e^-\right)}}
\newc{\brtauthreemu}{\ensuremath{\BR\left(\tau^{\pm}\rightarrow \mu^{\pm}\mu^+\mu^-\right)}}

\newcommand*{\reffig}[1]{Fig.~\ref{#1}}
 
        \newcommand*{\refeq}[1]{Eq.~(\ref{#1})}
     \newcommand*{\refsec}[1]{Sec.~\ref{#1}}

\newcommand*{\mstopone}{\ensuremath{m_{\tilde{t}_1}}}
\newcommand*{\mstoptwo}{\ensuremath{m_{\tilde{t}_2}}}

\let\oldcite\cite
\renewcommand*{\cite}{~\oldcite}

\newcommand*{\hl}{\ensuremath{h}}

\newc{\glzmu}{\ensuremath{{g^{Z\mu \mu}_{L}}}}
\newc{\grzmu}{\ensuremath{{g^{Z\mu \mu}_{R}}}}
\newc{\glwmu}{\ensuremath{{g^{W\mu \nu_\mu}_{L}}}}
\newc{\grwmu}{\ensuremath{{g^{W\mu \nu_\mu}_{R}}}}
\newc{\glzmuSM}{\ensuremath{{g^{Z\mu \mu}_{L,\textrm{SM}}}}}
\newc{\grzmuSM}{\ensuremath{{g^{Z\mu \mu}_{R,\textrm{SM}}}}}
\newc{\glwmuSM}{\ensuremath{{g^{W\mu \nu_\mu}_{L,\textrm{SM}}}}}
\newc{\grwmuSM}{\ensuremath{{g^{W\mu \nu_\mu}_{R,\textrm{SM}}}}}

\restylefloat{figure}

\begin{document}

\title{\LARGE {\bf The discreet charm of higgsino dark matter -- a pocket review}}

\author{\\ Kamila Kowalska\footnote{\url{kamila.kowalska@ncbj.gov.pl}}\, and Enrico Maria Sessolo\footnote{\url{enrico.sessolo@ncbj.gov.pl}} \\[2ex]
\small {\em National Centre for Nuclear Research,}\\
\small {\em Ho{\. z}a 69, 00-681 Warsaw, Poland }\\
}
%
\date{}
\maketitle
\thispagestyle{fancy}
\begin{abstract}
We give a brief review of the current constraints and prospects for detection of higgsino dark matter 
in low-scale supersymmetry. 
In the first part we argue, after performing a survey of all potential dark matter 
particles in the MSSM, that the (nearly) pure higgsino is the only candidate emerging virtually unscathed 
from the wealth of observational data of recent years. In doing so by virtue of its gauge quantum numbers and electroweak symmetry 
breaking only, it maintains at the same time a relatively high degree of model-independence. 
In the second part we properly review the
prospects for detection of a higgsino-like neutralino in direct underground dark matter searches, collider searches, 
and indirect astrophysical signals.
We provide estimates for the typical scale of the superpartners and fine tuning 
in the context of traditional scenarios where the breaking of supersymmetry is mediated at about the scale of Grand Unification and where 
strong expectations for a timely detection of higgsinos in 
underground detectors are closely related to the measured 125\gev\ mass of the Higgs boson at the LHC.   
\end{abstract}
\newpage 

\tableofcontents

\setcounter{footnote}{0}

\section{Introduction\label{sec:intro}}

From the particle physics point of view, the simplest, most popular, and arguably most robust mechanism leading
to the correct amount of cold dark matter (DM) in the early Universe is thermal freeze-out (see, e.g.,\cite{Kolb:1990vq,Gondolo:1990dk,Jungman:1995df,Dodelson:2003ft}). Briefly stated, one assumes 
that the DM consists of one or more matter species that were originally in thermal equilibrium with the Standard Model (SM) 
after the Big Bang and that, as the Universe expanded and cooled down, ``froze'' out of equilibrium when
their number density became too low for annihilation and creation processes to take place.

As is well known, in the context of the freeze-out mechanism the measurement of the relic abundance provided by 
WMAP and Planck, $\Omega_{\textrm{PL}}h^2=0.1188\pm0.0010$\cite{Komatsu:2010fb,Ade:2015xua}, 
implies a rather specific value for the thermally averaged annihilation 
cross section of the DM into SM particles: $\langle\sigv\rangle\approx 3\times 10^{-26}\,\textrm{cm}^3/\textrm{s}\approx 1\,\textrm{pb}$.    
Nevertheless, the thermal mechanism fails to provide any additional information on the nature of the DM itself since 
a cross section of that size can result from a discouraging wide 
range of DM mass values, spin quantum numbers, and DM-SM coupling strengths. Thus, in lack of more information, one 
has almost always to resort to some theoretical assumptions in order to narrow the search for DM down.

Since the 1990s, expectations about the scale of the new physics beyond the SM (BSM) 
have been driven by the theorists' discomfort with the hierarchy problem. 
This is the well-known fact that in a low-energy effective theory 
that includes one or more light fundamental scalars (as likely is the SM with a Higgs boson), 
one expects enormous quantum corrections to the scalar's mass from the physics in the UV 
(the Planck scale, in the absence of anything else).
Given the broad separation between the characteristic energies in play, this means that in order 
to get electroweak symmetry breaking (EWSB) one should fine tune the fundamental 
(unknown) Lagrangian parameters at the level -- again in the absence of anything lighter than the Planck scale --
of one part in $\sim 10^{28}$. 
Unless, of course, additional degrees of freedom were present, 
preferably close to the Higgs mass itself (say $\sim 100-1000\gev$).  

Remarkably, simply on dimensional grounds, if one of these expected TeV-scale BSM particles 
were to be the DM, its coupling to the SM extracted from the freeze-out mechanism would be of the size of the 
electroweak coupling constant,
$g\approx (16\pi m_{\textrm{DM}}^2\cdot1\,\textrm{pb})^{1/4}\approx 0.1-1$.
This fascinating coincidence, which, in light if its singling out specifically weakly interacting massive particles, or WIMPs, is known as the 
``WIMP miracle,'' maintains its attractiveness to these days, even if the LHC has failed to discover new particles below the scale of 
approximately 2\tev\cite{Atlas_LHC,CMS_LHC}.

Arguably the most complete and well motivated of the known BSM theories still 
remains low-scale supersymmetry (SUSY) (see, e.g.,\cite{Martin:1997ns}, for a popular review).
From the theoretical point of view, not only does SUSY provide possibly the most elegant solution to the hierarchy problem 
(if one allows for the possibility that, given the current LHC bounds, the theory might have to be amended to regain full naturalness); 
it also leads to a more precise UV unification of the gauge couplings than in the SM alone; 
it provides a solid rationale for the measured value of the Higgs boson and top quark masses and, by extension, for radiative EWSB.
From the phenomenological point of view, the Minimal Supersymmetric Standard Model (MSSM), 
contains all the necessary ingredients for successful baryogenesis and provides a framework for cosmic inflation.
It thus makes sense that, of all possible candidates for WIMPs, through the years a lot of attention was 
dedicated to the particles of the MSSM. 

In this review we give a compact summary of the subject of DM in the traditional MSSM. 
After briefly surveying the particles with the potential of providing a good DM candidate, we argue that the nearly pure higgsino neutralino 
survives to these days as perhaps the only one that is not in substantial tension with any phenomenological constraint. 
Interestingly, it does so in a relatively model-independent way, 
without the need of resorting to narrow or secluded regions of the parameter space.
We will thus review the higgsino's prospects for detection in direct underground DM searches, 
indirect searches for DM in gamma-ray and neutrino telescopes, and at the LHC. 
Incidentally we will show that, in those models where SUSY breaking is transmitted to the visible sector 
at the scale of Grand Unification (GUT), 
the detection prospects of higgsino DM become tightly bound to the typical mass of the sfermions in the spectrum 
and, as a direct consequence, to the size of the Higgs boson mass.  

In recent months several comprehensive reviews on the status of WIMP dark matter have appeared in the literature\cite{Gelmini:2016emn,Arcadi:2017kky,Plehn:2017fdg,Roszkowski:2017nbc}, 
one of which, co-authored by one of us, dedicated a full chapter to the MSSM neutralino
with particular attention to the detection prospects of a $\sim1\tev$ higgsino.  
While that work is broader in scope, casting light on the experimental opportunities provided by 
neutralinos in the context of the wider picture of thermal DM models, DM constraints, and existing experimental anomalies, 
we concentrate here instead on the specific physical characteristics of higgsinos, underlining 
what we believe makes them currently stand out as the most interesting elements in the DM panorama of the MSSM.
In this we are not dissimilar, perhaps, to recently appeared studies in the same tone\cite{Baer:2016ucr,Krall:2017xij}.
 
The structure of the review is as follows. In \refsec{sec:mssmdm} we recall
the particles of the MSSM that can provide a good DM candidate, classifying them according to their transformation 
properties under the SM gauge symmetry group. In \refsec{sec:pheno} we single out the 
higgsino as the most promising candidate of the list and review its detection prospects in different and complementary experimental 
venues. We dedicate an additional subsection to the calculation of typical fine tuning and expectations for the scale of the superpartners 
in models constrained at the GUT scale. 
We summarize the main treated points and conclude in \refsec{sec:sum}.

\section{Dark matter in the MSSM}\label{sec:mssmdm}

One of the features making the MSSM very attractive from a phenomenological point of view is 
that its gauge symmetry structure originates directly from the 
supersymmetrization of the SM itself. As such, the fundamental gauge symmetry is SU(3)$\times$SU(2)$\times$U(1),
and the dimensionless couplings are of the strong, electroweak, or SM Yukawa type.       

One of the consequences is that a potentially 
viable DM particle is also expected to interact with SM-like strength. 
Since cosmological observations have long excluded the possibility of DM 
particles being charged under color\cite{PhysRevD.41.3594} and, on the other hand, the DM is by definition ``dark,'' or practically 
electrically neutral\cite{Smith:1979rz,Jungman:1995df}, 
one is led to conclude that all viable DM candidates in the MSSM must be classifiable on the basis only of the SU(2) representation they belong to. Moreover, the available representations are limited  
to those that can be found in the SM: SU(2) singlets, doublets, and the adjoint.  

Before we proceed to briefly review these three groups individually, we 
remind the reader that in order to make the lightest SUSY particle (LSP) stable on cosmological time scales, one introduces
in the MSSM an additional discrete symmetry, R-parity\cite{Farrar:1978xj,Dimopoulos:1981zb,Weinberg:1981wj,Sakai:1981pk,Dimopoulos:1981dw}, 
under which only the superpartners of the SM fermions, gauge bosons, and any Higgs scalar field are odd. The origin 
of R-parity is still an active subject of research, and addressing the issue goes beyond the scope of the present review. 
We just point out that R-parity violation is strongly constrained phenomenologically, by the proton decay rate and electroweak precision  measurements\cite{Barbier:2004ez}. 

The only particles of the MSSM that are electrically and color-neutral are the neutrinos, their scalar superpartners, 
called \textit{sneutrinos}, and, finally, the \textit{neutralinos}.
Neutralinos, $\chi_{i=1,..,4}$, are Majorana fermion mass eigenstates emerging, after EWSB, 
from the diagonalization of the mass matrix of four electrically and color-neutral
SUSY states (see\cite{Ellis:1983ew,Griest:1988ma,Griest:1988yr,PhysRevD.41.3565} for early studies and\cite{Jungman:1995df} for a comprehensive, classic review). 
Two of these particles are \textit{gauginos}, fermionic superpartners of the SM gauge bosons. The \textit{bino}, $\tilde{B}$, 
in particular, is the partner of the U(1) gauge boson, while the neutral \textit{wino}, $\tilde{W}$, 
is the partner of the SU(2) gauge boson $W_3$. 
The other two states are neutral \textit{higgsinos}, $\tilde{H}_u$ and $\tilde{H}_d$, 
which belong to a vector-like pair of Higgs doublet superfields. 
If the lightest neutralino, hereafter indicated simply with $\chi$, is the LSP it can be the DM particle. 

At the tree level, the neutralino mass matrix takes the well-known form
\bea\label{neutmatr} \mathbf{M_\chi}=
  \begin{bmatrix}
    M_1 & 0 & -\frac{g' v_d}{\sqrt{2}} & \frac{g' v_u}{\sqrt{2}} \\
    0 & M_2 & \frac{g v_d}{\sqrt{2}} & -\frac{g v_u}{\sqrt{2}} \\
   -\frac{g' v_d}{\sqrt{2}} & \frac{g v_d}{\sqrt{2}} & 0 & -\mu \\
   \frac{g' v_u}{\sqrt{2}} & -\frac{g v_u}{\sqrt{2}} & -\mu & 0 
 \end{bmatrix}, \eea  
where $g$ and $g'$ are SU(2) and U(1) gauge couplings, respectively, $v_u$ and $v_d$ are the vacuum expectation values (vev) 
of the neutral components of the scalar Higgs doublets, $M_1$ and $M_2$ are the soft SUSY-breaking bare masses of the 
bino and wino, respectively, and $\mu$ is the vector-like mass parameter of the Higgs doublet superfields.

In the remainder of this section we give an overview of the mentioned DM candidates of the MSSM, 
highlighting the strongest phenomenological constraints that can be applied in each case. 
We will not, however, discuss the neutrinos. It has been long known\cite{Tremaine:1979we,White:1984yj} 
that the SM neutrinos do not provide, on their own, a viable candidate for cold DM. 
Their mass is $\mathcal{O}(<\textrm{eV})$, so that they are relativistic at the time of decoupling 
and therefore incur strong constraints from structure formation\cite{Abazajian:2005xn,dePutter:2012sh,Lukash:2012tq}. 
On the other hand, heavy right-handed neutrinos, whose existence might be postulated on the ground of the observed neutrino masses, 
and could provide a naturally expected extension of the traditional MSSM, 
also do not provide a good candidate for DM because 
they are not protected by R-parity and therefore not stable over cosmological scales in most scenarios. 

\subsection{SU(2) singlets\label{sec:singlet}}

\textbf{(Nearly) pure bino.} The first SU(2) singlet DM candidate we present is the bino.
Because of EWSB, a pure bino state does not exist in the MSSM, but the lightest neutralino behaves
like a pure bino to a very good approximation,
after the diagonalization of $\mathbf{M_\chi}$, if $|M_1|\ll M_2, \mu$.

The interactions of the bino-like neutralino 
with the SM fields are easily found by directly supersymmetrizing the SM gauge-fermion-fermion interaction 
and applying the R-parity conservation constraint. The resulting vertex takes the form bino-sfermion-fermion, 
$\mathcal{L}\supset -X_L \tilde{f}_L \bar{\chi} P_L f-X_R \tilde{f}_R \bar{\chi} P_R f$, 
where tree-level couplings, $X_{L,R}=\sqrt{2}\,g'\,Y_{L,R}$, are expressed in terms of the 
hypercharge assignment $Y_{L,R}$ of the fermion Weyl spinors.

\begin{figure}[t]
\centering
\subfloat[]{
\label{fig:a}
\includegraphics[width=0.25\textwidth]{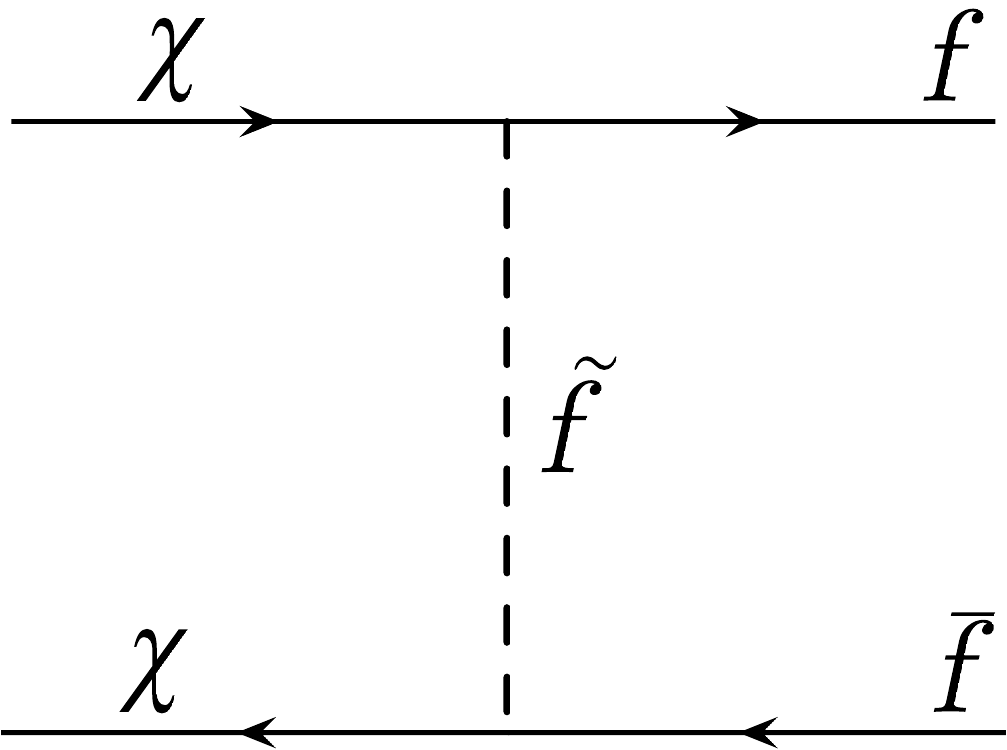}
}
\hspace{1cm}
\subfloat[]{
\label{fig:b}
\includegraphics[width=0.255\textwidth]{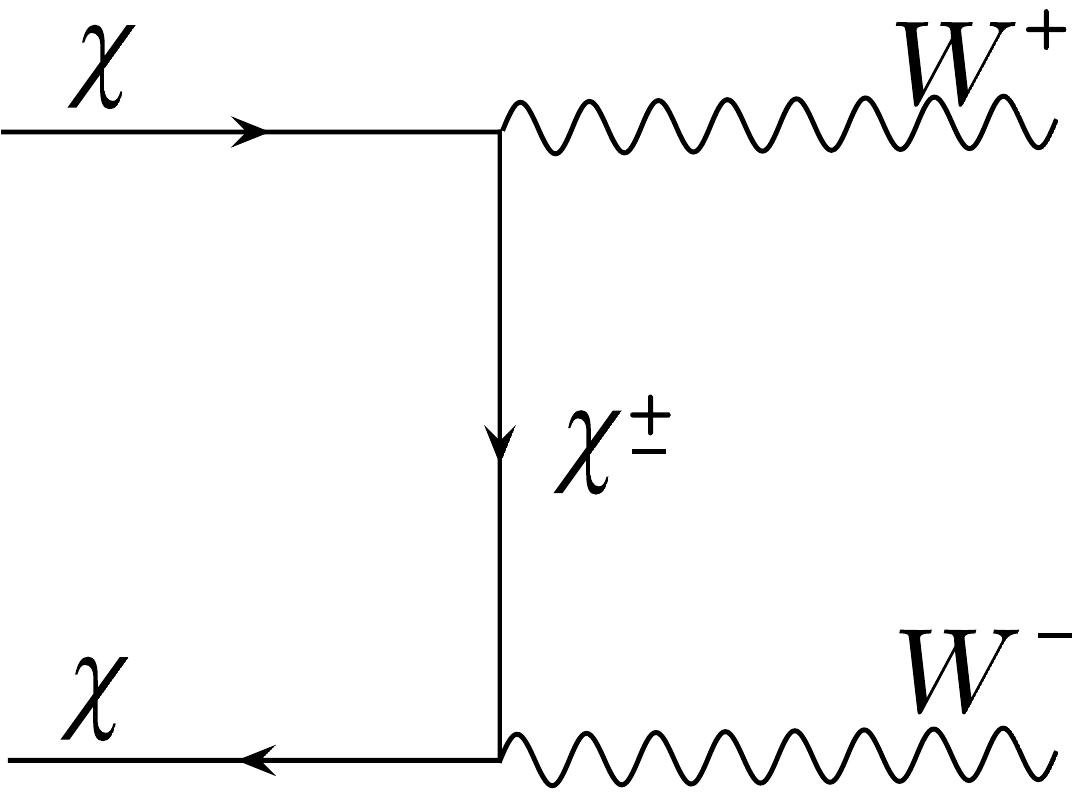}
}
\hspace{1cm}
\subfloat[]{
\label{fig:c}
\includegraphics[width=0.26\textwidth]{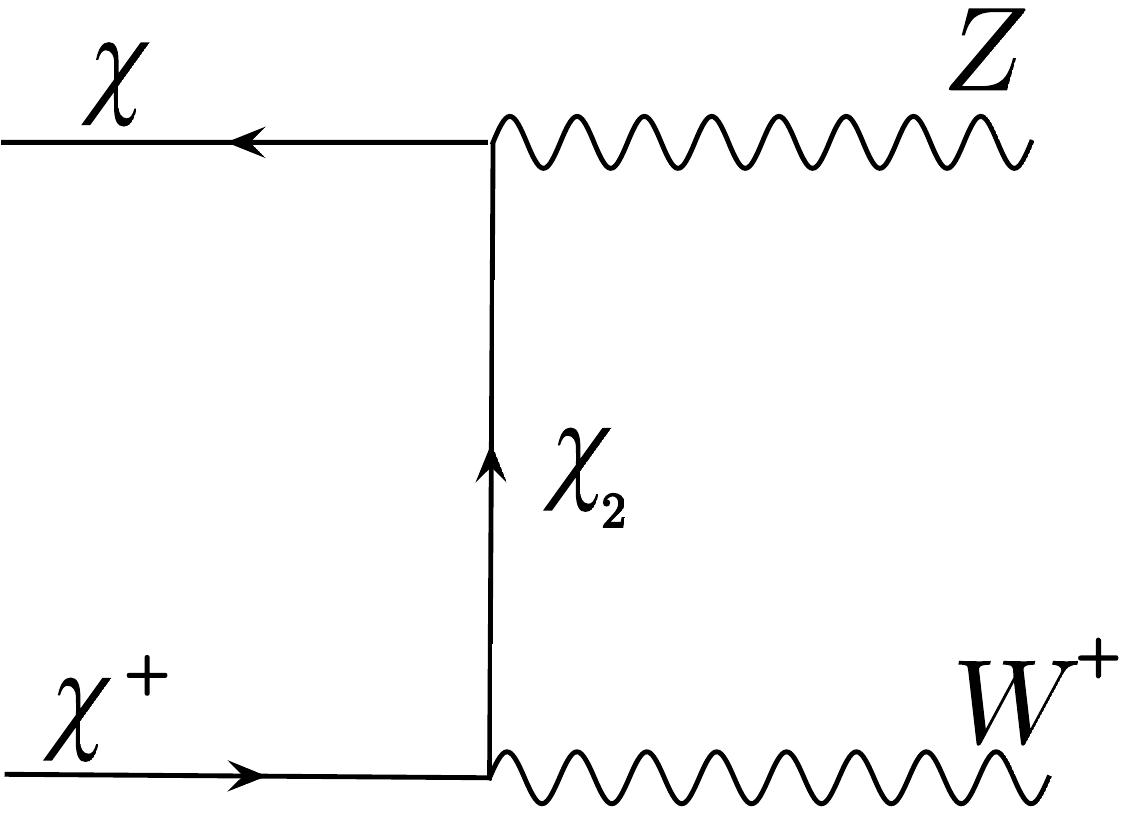}
}
\caption{(a) The dominant early-Universe annihilation channel for a nearly pure bino-like neutralino. (b), (c) 
Examples of annihilation and co-annihilation tree-level channels into gauge bosons for a predominantly higgsino-like neutralino.}
\label{fig:dmrelic}
\end{figure}

The pair-annihilation of bino-like neutralinos in the early Universe proceeds at the leading order through the 
$t$-channel diagram shown in \reffig{fig:dmrelic}(a). The region of the MSSM parameter space where $\abund\approx 0.12$ is
obtained in this way is historically known as the \textit{bulk}\cite{Drees:1992am,Baer:1995nc}. 
One can calculate the thermal cross section for binos,
given approximately by\cite{ArkaniHamed:2006mb}
\be\label{bulksigv}
\langle\sigv\rangle_{\tilde{B}} \approx \sum_{\tilde{f}}\frac{g'^4 Y_{\tilde{f}}^4}{2\pi}\,
\frac{\mchi^2\left(m_{\tilde{f}}^4+\mchi^4\right)}{\left(m_{\tilde{f}}^2+\mchi^2\right)^4}
\left(\frac{T_F}{m_{\chi}}\right)\,,
\ee
in terms of the neutralino (bino) mass, \mchi, 
sfermions' mass $m_{\tilde{f}}$, hypercharge $Y_{\tilde{f}}$, and freeze-out temperature $T_F$,
which parameterizes the dependence on velocity of the $p$-wave cross section, and is set here approximately at 
$T_F \approx (0.04-0.05)\,\mchi$.

The bulk has been long known to be strongly constrained by direct SUSY searches at colliders. 
To give a semi-quantitative estimate of these constraints, let us assume   
that only selectrons and smuons belong to the light SUSY spectrum, a 
reasonable ansatz in light of the strong LHC bounds on particles with color\cite{Sirunyan:2017kqq,Aaboud:2017aeu,Aaboud:2017vwy}. Assuming all four left- and right-handed slepton states have the same mass,
and inserting $Y_{\tilde{f}_L}=-1/2$, $Y_{\tilde{f}_R}=-1$ in \refeq{bulksigv}
one finds that the cross section is typically much smaller than 
$\sim 1\,\textrm{pb}$, except in the range $\mchi< m_{\tilde{f}}\lesssim 100\gev$.
A charged slepton mass of this size has been long excluded by direct searches at LEP\cite{Patrignani:2016xqp}.  

If, instead of selectrons and smuons, the light sfermions happen to be staus, 
the parameter space opens up a little, 
$m_{\tilde{\tau}_1}\lesssim 150\gev$ for $\mchi\approx 50\gev$,  
due to the non-negligible mixing between left and right chiral slepton  
states, which introduces an $s$-wave component to the annihilation cross section (see, e.g.,\cite{Fukushima:2014yia}). 
Nevertheless, LHC bounds on electroweak production\cite{Aad:2015eda}, implying $m_{\tilde{\tau}_1}\gsim 109\gev$, 
are by now becoming strongly constraining for these scenarios too, which will be probed even more deeply soon\cite{ATL-PHYS-PUB-2016-021}.
Finally, as we have mentioned, SUSY parameter space where bulk sfermions are charged under color is strongly 
excluded by LHC direct searches.

A way to evade the strong collider bounds is provided, if the bino-like neutralino 
and some other sparticles (sfermions $\tilde{f}$ or other gauginos) 
are nearly degenerate in mass, by the mechanism of co-annihilation\cite{Griest:1990kh,Ellis:1998kh,Ellis:1999mm}. 
In this case the cross section of \refeq{bulksigv} should be
replaced by an effective quantity that takes into account the thermal average of all annihilations and co-annihilations of the kind
$\chi\chi,\chi\tilde{f},\tilde{f}\tilde{f}\rightarrow \textrm{SM}\,\textrm{SM}$, some of which are likely to be much more efficient 
than $\chi\chi\rightarrow \textrm{SM}\,\textrm{SM}$ alone. 

However, without any guidance from the theory in the UV,
co-annihilation of the bino with other sparticles can only be achieved
in narrow slices of the parameter space, which require some tuning of the initial parameters to engineer the desired 
coincidence of neutralino and sfermion mass. And in models that are instead 
defined in terms of a limited number of free parameters in the UV, like the CMSSM\cite{Kane:1993td}, 
in which slepton or stop co-annihilation with the bino can occur naturally for particular choices of the initial conditions, 
the preferred regions of the parameter space are incurring increasingly strong limits from direct LHC 
searches\cite{Roszkowski:2014wqa,Bechtle:2015nua,Han:2016gvr,Athron:2017qdc,Roszkowski:2017nbc}. Besides, 
with gaugino universality at the GUT scale it is a struggle to fully accommodate the measured value of the Higgs mass 
at the LHC\cite{Roszkowski:2014wqa,Ellis:2018jyl} 
(this problem is resolved if the gluino mass is a free parameter, e.g.,\cite{Akula:2013ioa}).
Thus, even if co-annihilation of the bino with other sparticles can still lead to viable regions of the parameter space in the most generic 
parametrizations of the MSSM\cite{Roszkowski:2014iqa}, 
it is also perhaps not exceedingly attractive from a natural point of view.
\bigskip

\noindent \textbf{R sneutrino.} The second SU(2) singlet DM candidate of the MSSM is the scalar ``right-handed'' sneutrino. 
The right-handed sneutrino does not properly belong to the MSSM, which in its original formulation 
features massless neutrinos, but naturally emerges in SM extensions with right-handed neutrinos, 
which can give rise to the neutrino mass via small Yukawa couplings (if the right-handed neutrino is Dirac), 
or through the see-saw mechanism (if the right-handed neutrino is Majorana, see, e.g.,\cite{Mohapatra:1999em} and references therein). 

The phenomenology of right-handed sneutrinos as DM, however interesting, is very model-dependent. 
In traditional see-saw models with large-scale Majorana mass the right-handed sneutrino is too heavy to be the DM.
On the other hand, for a sneutrino of the ``Dirac'' type, or, in alternative, 
Majorana but such that the bare mass is of the order of the superpartners' mass\cite{ArkaniHamed:2000bq,Borzumati:2000mc},
the only really model-independent vertex with the SM involves a very 
small Yukawa coupling $\mathcal{L}\supset -y_{\nu_R}\,\bar{e}_L \tilde{H}_u^{\pm}\,\tilde{\nu}_R 
-y_{\nu_R}\,\bar{\nu}_L \tilde{H}_u^{0}\,\tilde{\nu}_R$. Thus, the induced 
$t$-channel processes similar to \reffig{fig:dmrelic}(a), with sneutrinos (charginos) in place of neutralinos (sfermions), 
and a tiny coupling constant, are not strong enough to get the correct \abund.  

On the other hand, the correct relic density can certainly be obtained thanks to the mixing with the left-handed sneutrino, 
and SUSY breaking can generate $A$-terms of the order of the SUSY scale, which provide large couplings to the SM Higgs boson. 
The phenomenology of these cases can be very rich and exceeds the scope of this review. 
We direct the reader to the vast literature on sneutrino DM for further details (see, e.g.,\cite{TuckerSmith:2001hy,TuckerSmith:2004jv,Asaka:2005cn,Arina:2007tm}, 
for early studies and bounds, and\cite{Arina:2015uea} for a recent LHC analysis).

\subsection{SU(2) doublets\label{sec:doublet}}

We have seen that singlet DM candidates in the MSSM are accompanied by some
uncomfortable features: they are either strongly constrained by collider bounds, are only viable in fine-tuned regions of the 
parameter space, or present a phenomenology that is highly model-dependent. 
We therefore move on to reviewing the next set of candidates, the SU(2) doublets.\smallskip

\noindent \textbf{(Nearly) pure higgsino.} The most popular SU(2) doublet DM candidate, 
and the one that appears to us most attractive from a phenomenological point of view,
is the higgsino, which is the main subject of this review.
As was the case for the bino, there is no pure higgsino state after EWSB, but
one obtains an almost pure higgsino-like neutralino by diagonalizing $\mathbf{M_\chi}$ in \refeq{neutmatr} in the limit $|\mu|\ll M_1, M_2$.

As supersymmetry assigns a Weyl spinor to each complex state in the scalar Higgs doublets 
one counts four physical higgsino states, which, after EWSB, give rise to
two Majorana neutralinos, $\chi_1$ (or $\chi$) and $\chi_2$, and a Dirac chargino, $\chi^{\pm}$. 
When $|\mu|\ll M_1\approx M_2$, 
the tree-level mass splitting between the two higgsino-like neutralinos is 
of approximately the size of $m_Z^2/M_{1,2}$\cite{Martin:1997ns},
and the splitting between the higgsino-like chargino and the lightest neutralino is approximately half of that. 
Moreover, radiative corrections also induce a non-negligible and irreducible mass splitting ($\sim 100$s~MeV) 
between the charged and neutral states 
(see, e.g.,\cite{Drees:1996pk,Nagata:2014wma}).

To correctly compute the thermally-averaged effective cross section that yields the DM relic abundance,
one must take into account all possible annihilations and co-annihilations of higgsino states. 
For \mchi\ above the $W$ threshold the dominant final state is into $W$ and $Z$ bosons (Figs.~\ref{fig:dmrelic}(b) and \ref{fig:dmrelic}(c) 
give examples of possible diagrams for this processes), 
to which higgsino-like neutralinos and charginos couple through the electroweak charged and neutral currents\cite{Jungman:1995df},
\be
\mathcal{L}\supset \left(-\frac{g}{2}\,W^+_{\mu}\bar{\chi}\gamma^{\mu}\chi^- -\frac{g}{4\cos\theta_W}\,Z_{\mu}\bar{\chi}_1\gamma^{\mu}\chi_2 +\textrm{h.c.}\right)-\frac{g}{2\cos\theta_W}\,Z_{\mu}\bar{\chi}^+\gamma^{\mu}\left(1 -2\sin^2\theta_W\right) \chi^-.
\ee

The effective cross section  can be obtained at the leading order 
in the limit of all four states being degenerate (see, e.g.,\cite{ArkaniHamed:2006mb}):   
\be\label{higgsinosigv}
\langle\sigv \rangle_{\tilde{H}}^{(\textrm{eff})} \approx \frac{21\,g^4+3\,g^2 g'^2+11\,g'^2}{512\,\pi\,\mchi^2}\,.
\ee

For heavy, very pure higgsinos, one should include in the calculation of $\langle\sigv \rangle_{\tilde{H}}^{(\textrm{eff})}$ 
corrections due to the Sommerfeld enhancement, a well known non-perturbative effect originating from the fact that if a DM particle is 
much heavier than the electroweak gauge bosons and relatively slow, the weak force becomes effectively long-range and the impact of 
the non-relativistic potential on the interaction cross section becomes significant\cite{Hisano:2002fk,Hisano:2004ds}. However, 
in the case of the higgsino the splitting between its charged and neutral components is almost always
large enough to effectively wash out substantial non-perturbative effects originating from the 
resummation of ladder diagrams\cite{Hisano:2006nn,Cirelli:2007xd,Hryczuk:2010zi}, so that in a first 
approximation \refeq{higgsinosigv} provides a fairly accurate estimate of $\langle\sigv \rangle_{\tilde{H}}^{(\textrm{eff})}$.

One can see that the cross section is typically much larger than $\sim 1\,\textrm{pb}$, unless
 $\mchi\approx 1\tev$ (the precise numerical value is more about 1.1\tev, as we shall see). 
Thus, a $\sim 1\tev$ higgsino is on its own a good candidate for the DM in the Universe\cite{Profumo:2004at}, 
while a higgsino much lighter than 1\tev\
requires one to assume the existence of an additional DM component
(e.g., axion\cite{Baer:2011hx,Baer:2011uz}), needed to get $\abund\approx 0.12$.

As we shall see in the next sections, a $\sim 1\tev$ higgsino is generally associated with a large
SUSY-breaking scale, and for this reason it is not currently very constrained from a phenomenological point of view.
However, its characteristic properties can give us hope for a timely detection in direct and indirect DM searches and even, 
if $\mchi\ll 1\tev$, in collider searches.
\bigskip

\noindent \textbf{L sneutrino.} We conclude this subsection by reviewing the properties of the only other SU(2) doublet DM candidate in the MSSM:
the ``left-handed'' sneutrino, scalar superpartner of the SM left-handed neutrino. 

The left-handed sneutrino is a complex scalar field with SU(2)$\times$U(1) quantum numbers equal to the higgsino's. 
Like the higgsino, it has charged and neutral current couplings to the $W$ and $Z$ bosons,
$\mathcal{L}\sim -ig/\sqrt{2}\left(W^+_{\mu}\tilde{\nu}^{\ast}_L \partial^{\mu}\tilde{e}^-_L
+W^-_{\mu}\tilde{e}^+_L \partial^{\mu}\tilde{\nu}_L\right)-ig/(2\cos\theta_W)\,Z_{\mu}\,\tilde{\nu}^{\ast}_L \partial^{\mu}\tilde{\nu}_L$\,. 
The mass splitting of the charged and neutral components of the SU(2) doublet is, however, 
much larger for sneutrinos/sleptons than for higgsinos, being generated 
through hypercharge D-term contributions\cite{Martin:1997ns}:
$m^2_{\tilde{e}_L}-m^2_{\tilde{\nu}_L}\approx -m_W^2\cos 2\beta$, where $\tanb\equiv v_u/v_d$. 
Thus, one should resist the temptation of interpreting \refeq{higgsinosigv} as an
accurate estimate of the effective cross section for sneutrinos too, 
since the co-annihilation of charged and neutral states becomes somewhat 
less efficient. It turns out\cite{Arina:2007tm} that the mass required to produce 
$\langle\sigv \rangle_{\tilde{\nu}_L}^{(\textrm{eff})}\approx 1\,\textrm{pb}$ is about $m_{\tilde{\nu}_L}\approx 600-700\gev$.
Sneutrinos lighter than that imply the existence of an additional component of DM.
 
A very important constraint on left-handed sneutrinos as DM arises because they, unlike the Majorana higgsino-like neutralinos, 
are not their own antiparticle, so that
their elastic scattering with nuclei in direct detection experiments proceeds also through $t$-channel exchange of a $Z$ boson. 
By virtue of the sneutrino's neutral current coupling, the spin-independent cross section is approximately given 
by a Fermi-like contact interaction, 
$\sigsip\approx \mu_{\textrm{red}}^2 G_F^2/8\pi\approx 10^{-3}\,\textrm{pb}=10^{-39}\,\textrm{cm}^2$,
where reduced mass $\mu_{\textrm{red}}\approx m_p$ for $m_{\tilde{\nu}_L}\gg m_p$. Cross sections of this size  
have been long excluded in underground detector searches\cite{Falk:1994es,Hall:1997ah}.

\subsection{SU(2) adjoint triplet\label{sec:triplet}}    
 
\textbf{(Nearly) pure wino.} The only SU(2) triplet DM candidate in the MSSM is the wino-like neutralino, dominated by 
the fermionic superpartner of the $W_3$ weak gauge boson. 
The wino belongs to the adjoint representation of the gauge group (hypercharge $Y=0$) and the wino-like neutralino emerges, after EWSB, from the diagonalization of \refeq{neutmatr} in the limit $|M_2|\ll M_1, \mu$. 
One finds a Majorana neutralino, $\chi$, and a Dirac chargino, $\chi^{\pm}$, mass-degenerate at the tree level. 
In the context of UV complete models of SUSY-breaking, spectra with a light wino 
can arise, for example, in scenarios where SUSY breaking is transmitted via anomaly mediation\cite{Randall:1998uk,Giudice:1998xp}. 

If the wino LSP is heavier than the electroweak gauge bosons, its dominant final state channel for annihilation (and co-annihilation with charginos) in the early Universe is into $W$ (but not $Z$) boson final states, to which it couples as 
$\mathcal{L}\sim -g\,W^{\pm}_{\mu}\bar{\chi}\gamma^{\mu}\chi^{\mp}$.
The thermal annihilation cross-section is dominated by coannihilations of the three wino states, 
similarly to what happens for the doublet higgsinos. 
Annihilation into fermion–antifermion final states through a $t$-channel sfermion exchange, 
reminiscent of the bino bulk mechanism, has been instead long excluded by LEP limits on the charged slepton masses.
   
Unlike higgsinos, in the wino case mass splitting between the charged and neutral fermion component of the 
SU(2) multiplet is generated exclusively by radiative corrections, 
$\Delta M_{\widetilde{W}}=(g^2/4\pi)\,m_W \sin^2 (\theta_W/2)\approx 166\mev$\cite{Cirelli:2005uq}.
Note that the mass splitting is typically much smaller than for higgsinos, 
so that one cannot neglect the effects 
of the Sommerfeld resummation on the calculation of the thermal cross section.
When one includes the Sommerfeld enhancement numerically, the correct relic density is obtained
for $\mchi\approx 2.7-2.8\tev$\cite{Hisano:2006nn,Cirelli:2007xd,Hryczuk:2010zi}. For a lighter mass, winos do not saturate the relic abundance.

The Sommerfeld enhancement induces more dramatic modifications of the effective DM annihilation 
cross section when the average kinetic energy of the WIMP corresponds to speeds of the order of $10^{-3} c$, 
as in the present-day Universe.
This fact has led to the derivation of powerful indirect astrophysical constraints on the annihilation cross section of wino-like 
neutralinos\cite{Cohen:2013ama,Fan:2013faa,Hryczuk:2014hpa,Beneke:2016jpw,Cuoco:2017iax}.
By taking into account the effects of Sommerfeld-enhanced contributions to the annihilation of winos into 
mono-chromatic gamma rays, as well as bounds on the present-day cross section to $W^+W^-$ from diffuse gamma radiation 
from the Galactic Center and Dwarf Spheroidal satellite galaxies (dSphs), measured in terrestrial and space telescopes 
H.E.S.S.\cite{Abramowski:2013ax,Abdallah:2016ygi} and Fermi-LAT/MAGIC\cite{Ahnen:2016qkx}, and from cosmic ray (CR) antiproton data at AMS-02\cite{Cuoco:2017iax,Aguilar:2016kjl},
one can derive strong independent constraints (albeit affected by significant systematic uncertainties)
which steeply raise the stakes on the wino as a viable DM particle, 
especially in scenarios where it saturates the relic abundance. 

\subsection{Mixed cases\label{sec:mixed}}

The four neutralinos of the MSSM are all Majorana fermions that, after EWSB, 
remain neutral under $U(1)_{\textrm{em}}$ and color. 
In the absence of a well-separated hierarchy among $M_1$, $M_2$, and $\mu$, the lightest mass eigenstate 
will be an admixture of the SU(2) gauge multiplets discussed in Secs.~\ref{sec:singlet}-\ref{sec:triplet} but, unlike those cases, 
it will present properties that differ significantly from a pure gauge eigenstate.

When $|M_1|\approx |\mu|$ the neutralino is in a highly mixed bino/higgsino state. 
Mixed neutralinos of this kind (sometimes also called ``well-tempered''\cite{ArkaniHamed:2006mb}), 
originally observed in mSUGRA parameter space\cite{Chan:1997bi,Feng:1999zg,Feng:2000gh}
but that can arise under different boundary conditions (e.g.,\cite{Baer:2006te,Baer:2008ih}),
enjoyed some popularity, 
especially before the advent of the LHC, because they can easily lead to $\abund\approx 0.12$ for values of the $\mu$ 
parameter as low as few hundreds~GeV, which are favored to solve the hierarchy problem.
However, the rapid progress made in the bounds on the spin-independent cross section 
of the neutralino scattering off nuclei in direct WIMP detection searches, 
combined with a failure to directly observe scalar fermions and heavy Higgs bosons at the LHC, 
have rendered scenarios where the lightest neutralino is a rich admixture of gaugino and higgsino 
much less appealing if not excluded altogether (see, e.g.\cite{Badziak:2017the}, for a very recent update of the constraints 
on bino-higgsino, and\cite{Beneke:2016jpw} for wino-higgsino scenarios).

\begin{figure}[t]
\centering
\includegraphics[width=0.2\textwidth]{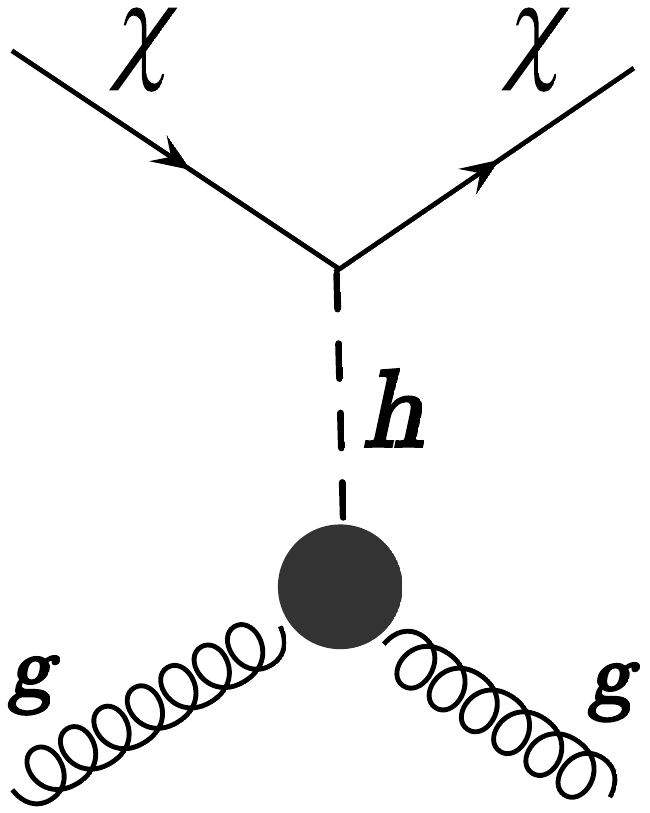}
\caption{The main interaction between the 
neutralino and heavy nuclei in underground detectors  in the limit of squarks and heavy Higgs bosons being much heavier than 
$\mhl=125\gev$ and in general outside of LHC reach.}
\label{fig:sigsip}
\end{figure}

To briefly set the issue on quantitative grounds, 
let us estimate the strength of the coupling with which neutralino admixtures of higgsino and gaugino 
contribute to the spin-independent cross section.
We recall that, in the limit of the squarks and heavy Higgs bosons being much heavier than 
$\mhl=125\gev$, which has become a reasonable assumption after the first two runs of the LHC, the main interaction between the 
neutralino and heavy nuclei in underground detectors proceeds as in \reffig{fig:sigsip},
via $t$-channel exchange of the 125\gev\ Higgs boson and an effective coupling to gluons through the heavy quark loops. 

As the neutralino LSP-Higgs-neutralino LSP tree-level vertex directly stems from applying the gauge covariant derivative on the Higgs doublets, 
it is non-zero only for a gaugino/higgsino admixture. 
For \tanb\ 
sufficiently large to ensure a predominantly SM-like Higgs boson,\footnote{$\tanb>3-4$ is a condition often fulfilled, 
for instance, in scenarios where EWSB is obtained radiatively via the renormalization group evolution
of soft SUSY-breaking parameters constrained at some high scale, as it prevents certain 
soft masses from running tachyonic at the low scale.}  
the coupling to the nucleon can thus be expressed entirely in terms of the higgsino fraction 
(or \textit{purity}), $f_h$, which depends 
on the elements of the unitary matrix, $N$, diagonalizing \refeq{neutmatr}.

If $\textrm{diag}[m_{\chi_1},m_{\chi_2},m_{\chi_3},m_{\chi_4}]=N\,\mathbf{M_\chi}N^{\dag}$, 
one can define $f_h\equiv |N_{13}|^2+|N_{14}|^2$ and express the coupling of interest as 
$\mathcal{L}\sim (g\sqrt{f_h\left(1-f_h\right)}/4)\bar{\chi}\chi h$\,.
Note, incidentally, that deriving an explicit form for the elements of matrix $N$ in terms of bare masses 
$M_1$, $M_2$, and $\mu$ is not a trivial task even at the tree level, 
and useful formulas in this regard can be found in several papers, for example\cite{ElKheishen:1992yv,Choi:2001ww,Choi:2004rf,Beylin:2008zz}.
By simple inspection of \refeq{neutmatr}, however, one can infer a rough approximation for the higgsino fraction in the limit
of nearly pure higgsinos,  $|\mu|\ll M_2\approx M_1$:
\be\label{hinofrac}
1-f_h\approx\frac{m_W^2}{(M_{1,2}-|\mu|)^2}\,.
\ee
Equation~(\ref{hinofrac}) becomes quite accurate for $f_h \gsim 0.999$.

The spin-independent cross section of the neutralino with protons (nucleons),
$\sigsip=\left(4\mu_{\textrm{red}}^2/\pi\right) \left|\mathcal{A}_p\right|^2$, can be parameterized for moderate-to-large \tanb\ 
simply as\cite{Jungman:1995df}
\be\label{hinosigsip}
\mathcal{A}_p(f_h)\approx a_{\textrm{eff}}\,\frac{f_{TG}}{9}\,\frac{m_p}{v}\,\frac{g\sqrt{f_h\left(1-f_h\right)}}{m_h^2}\,,
\ee   
in terms of the gluon fractional content of the proton, $f_{TG}$ (we use the default value for 
\texttt{micrOMEGAs~v4.3.1}\cite{Belanger:2013oya}, $f_{TG}=0.92$), and a phenomenological fudge factor,
$a_{\textrm{eff}}\approx 0.9-1$, which takes into account the dependence of $\mathcal{A}_p$ on twist-two operators\cite{Drees:1993bu} 
and higher-order loop corrections\cite{Hisano:2004pv}.

\begin{figure}[t]
\centering
\includegraphics[width=0.55\textwidth]{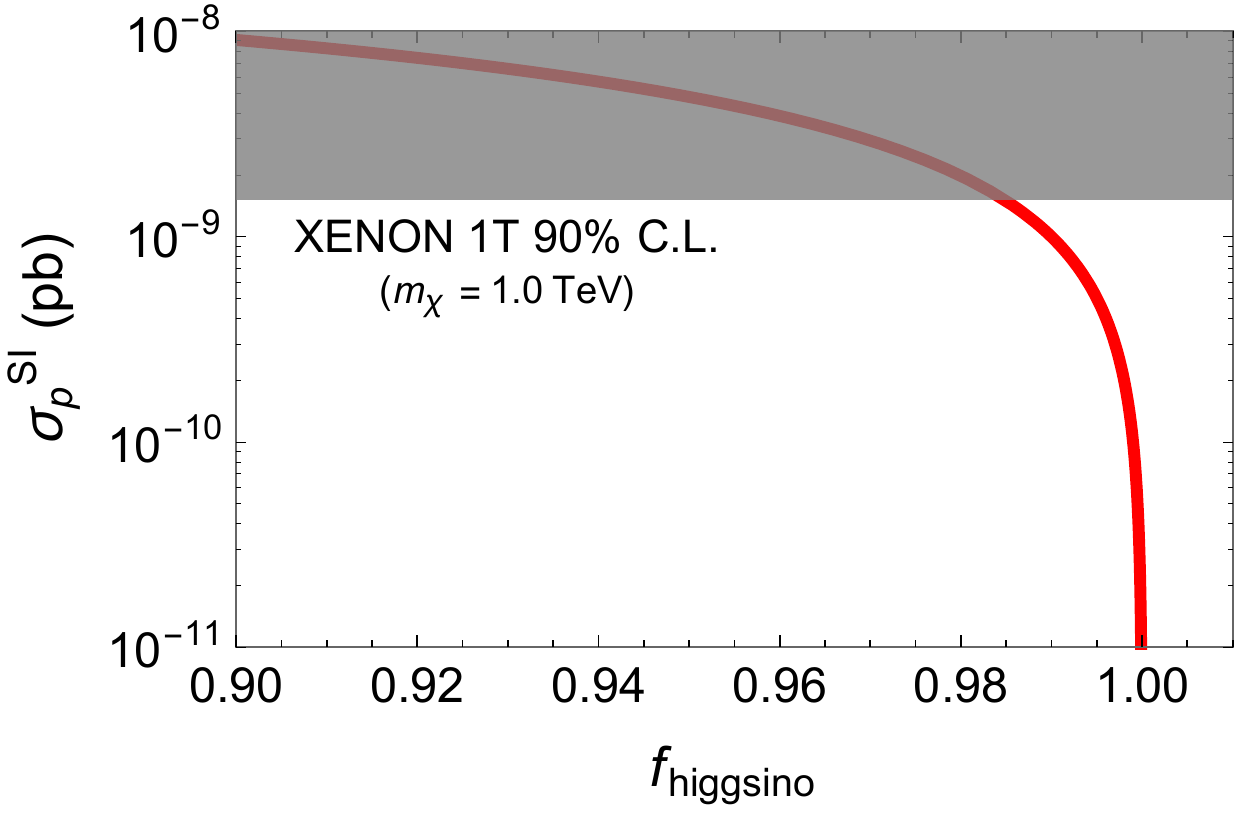}
\caption{The neutralino-proton spin-independent cross section, \sigsip,
for a typical case of predominantly higgsino-like neutralino DM with $\mchi=1.0\tev$
as a function of higgsino purity $f_{\textrm{higgsino}}$ ($\equiv f_h$).
}
\label{fig:purity_sigsip}
\end{figure}

We show in \reffig{fig:purity_sigsip} a plot of \sigsip\ 
as a function of purity $f_h$ for a $\mchi=1\tev$ neutralino
(to a first approximation the DM mass affects the cross section only through the reduced mass leading to
 $\mu_{\textrm{red}}\approx m_p$).
One can see that, for admixtures dominated by the higgsino fraction, 
the most recent XENON-1T 90\%~C.L. upper bound\cite{Aprile:2017iyp} on \sigsip\
enforces $f_h>98\%$, so that viable DM candidates ought to be very close to a pure higgsino state.

Since the purity of well-tempered higgsino-dominated neutralinos stays well below~90\% 
in those models attempting to provide a satisfactory solution to 
the hierarchy problem while saturating the relic abundance\cite{ArkaniHamed:2006mb}, 
we conclude that, barring increasingly narrow corners of the parameter space\cite{Badziak:2017the}, 
these scenarios have become very hard to rescue or justify in light of the most recent direct detection bounds.
\bigskip

To conclude this subsection, we finally recall that in cases where 
$|M_1|<|\mu|\lesssim 1-2\tev$, one obtains scenarios where the mixed neutralino is predominantly bino-like, 
but also acquires couplings that originate from its admixture with higgsino states, so that additional mechanisms for 
obtaining $\langle\sigv\rangle\approx1\,\textrm{pb}$ with respect to \refsec{sec:singlet} are possible.

These mechanisms, often called \textit{funnels}, 
involve resonant or close-to-resonant $s$-channel annihilation of two neutralino LSPs 
via a nearly on-shell mediator which could be the $Z$ boson (if $\mchi\approx m_Z/2$)\cite{Griest:1988ma}, 
the SM Higgs boson (if $\mchi= 60-65\gev$)\cite{Ellis:1989pg}, 
or one of the heavy Higgs bosons of the MSSM\cite{Drees:1992am}. 

Note that the $Z$-funnel parameter space 
is strongly constrained by the LHC. The coupling of the lightest neutralino to the $Z$ boson is due exclusively 
to the isospin neutral current, cf. \refsec{sec:doublet}, which means that in mixed 
bino-higgsino scenarios it is directly proportional to the 
higgsino fraction. As a consequence, $f_h$ cannot take excessively small values or, in other words, 
$\mu$ cannot be much larger than $M_1\approx m_Z/2$. 
The relative proximity of a mostly higgsino-like chargino and a mostly bino-like neutralino subjects this region of the 
parameter space to strong bounds from direct LHC multi-lepton searches\cite{Calibbi:2014lga}.

Light and heavy Higgs boson funnels are less constrained from direct LHC SUSY searches than the $Z$ funnel,
since the direct coupling to the lightest neutralino is dependent on $\sqrt{f_h}$ and the mediator can be quite heavy.
However, there exist complementary observables which can constrain these regions, 
like the branching ratio \brbsmumu\cite{Kowalska:2013hha} and direct 
searches for heavy Higgs bosons in the $\tau\tau$ channel\cite{Arbey:2013jla}. 
Moreover, as was the case for the co-annihilations of the bino,   
most phenomenological scenarios require \textit{ad hoc} arrangement of the parameters to obtain the right ratio of 
neutralino to scalar mass, although this is not necessarily the case for some parameter-space regions of 
GUT-constrained scenarios like the CMSSM, in which the renormalization group evolution (RGE) of soft masses from 
a handful of free parameters can lead more naturally to 
the right mass coincidence (see, e.g.,\cite{Lahanas:1999uy,Ellis:2001msa} for early studies).

\section{Phenomenology of higgsino dark matter}\label{sec:pheno}
 
The discussion of \refsec{sec:mssmdm} has led us to conclude that the sole DM candidate of the MSSM 
emerging almost unscathed from the wealth of observational data of recent years is the nearly pure higgsino. 
We therefore dedicate this section to the analysis of the prospects for detection of a higgsino-like neutralino in direct DM detection 
searches, collider searches, and indirect astrophysical signals, 
and spend a few words on alternative strategies in other experimental venues. We will also 
give some predictions for the scale of the superpartner particles in traditional models and briefly discuss the issue of fine tuning.

\subsection{Prospects for detection in direct and indirect searches}\label{sec:prosp}

\begin{figure}[t]
\centering
\subfloat[]{
\includegraphics[width=0.50\textwidth]{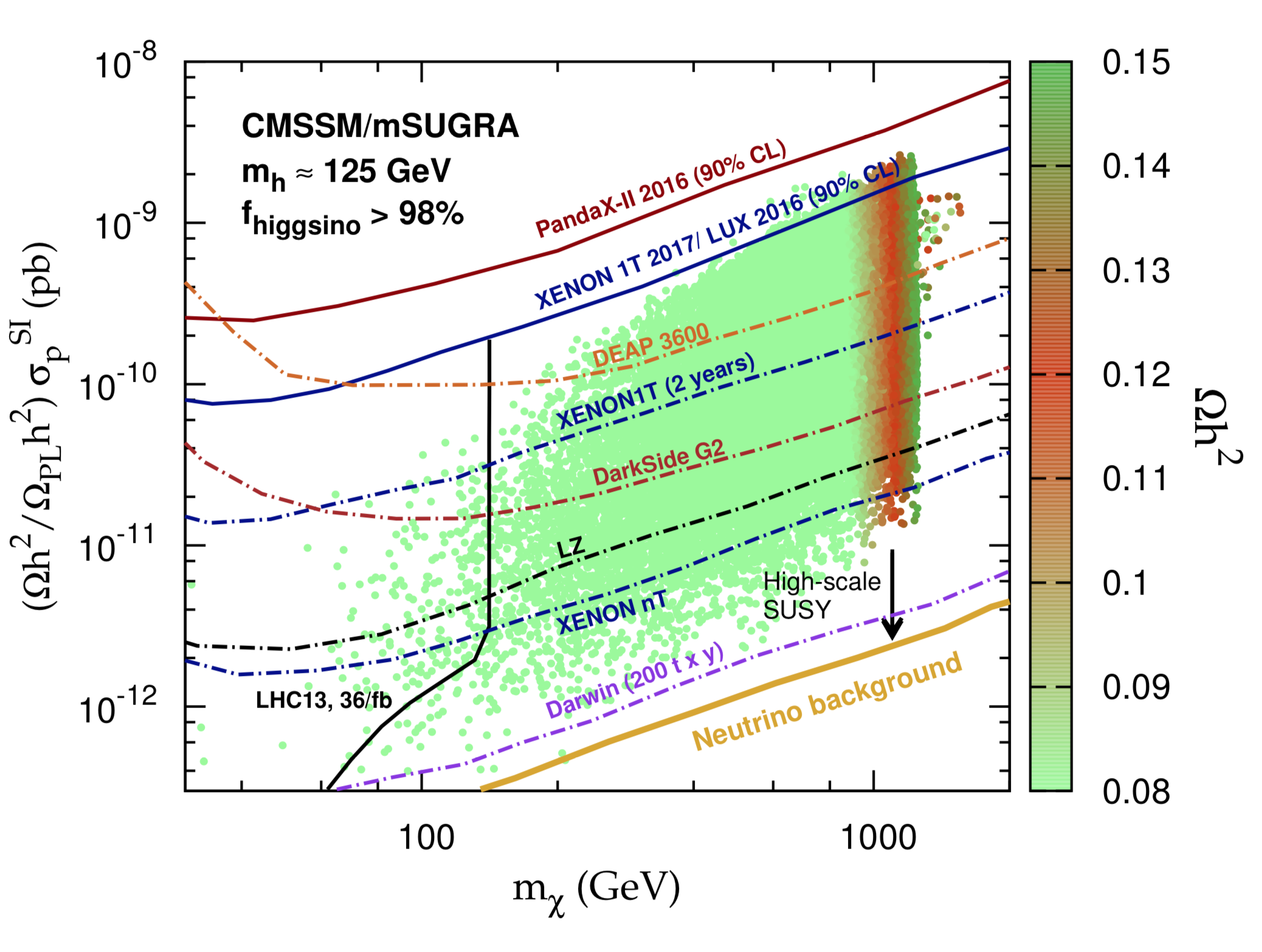}
}
\subfloat[]{
\includegraphics[width=0.50\textwidth]{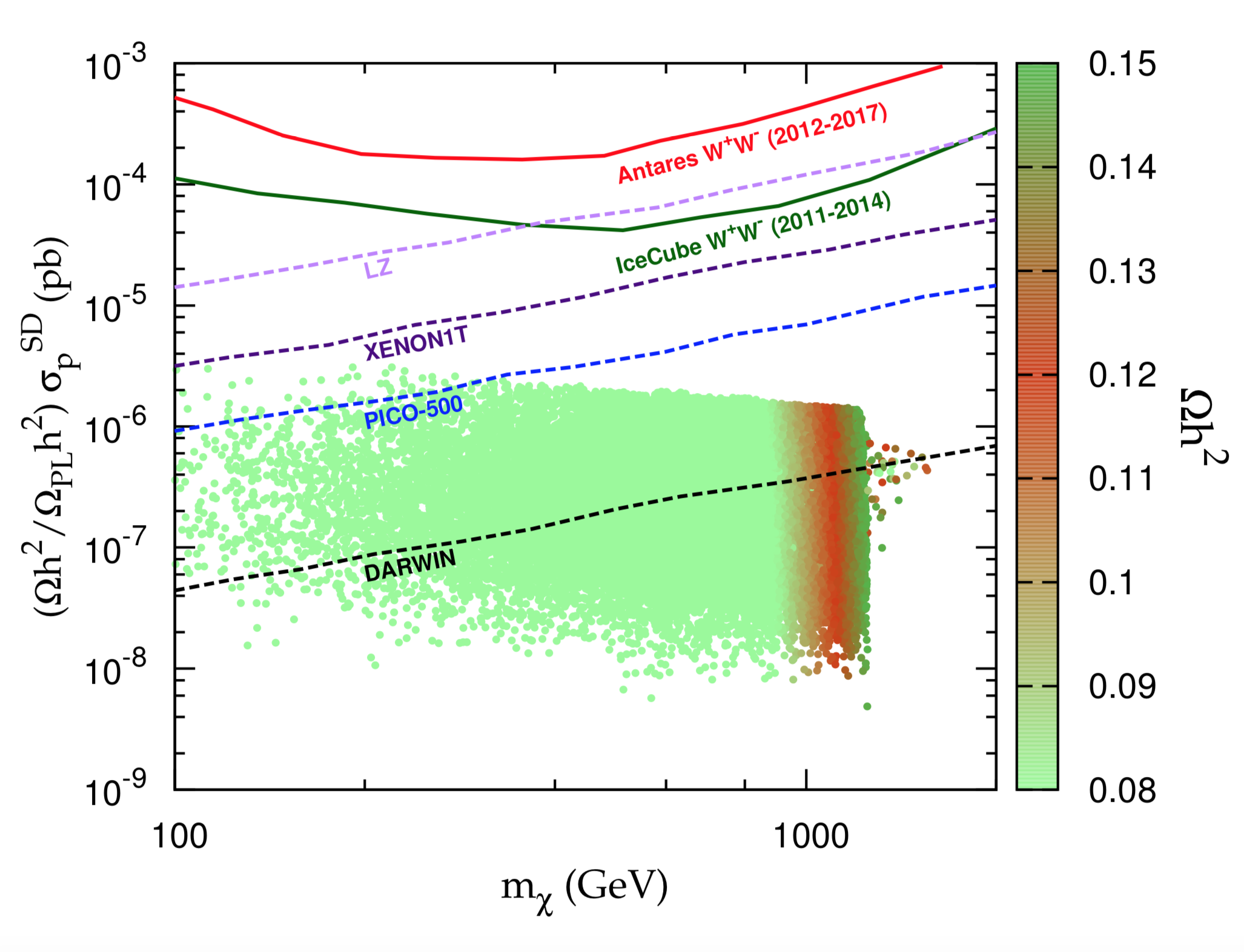}
}
\caption{(a) Spin-independent neutralino-nucleon cross section \sigsip\ rescaled by the relic abundance, 
as a function of neutralino mass \mchi, for a nearly pure higgsino with CMSSM/mSUGRA boundary conditions subject to $m_h\approx 125\gev$ and LHC Higgs bounds. Solid lines show the 90\%~C.L. upper bounds from PandaX-II\cite{Tan:2016zwf} (maroon) and XENON1T\cite{Aprile:2017iyp} (LUX\cite{Akerib:2016vxi}) (blue). Dot-dashed lines show the projected reach for DEAP-3600\cite{Amaudruz:2014nsa} (orange), XENON1T/nT\cite{Aprile:2015uzo} (blue), DarkSide G2\cite{Aalseth:2015mba} (maroon), LZ\cite{Szydagis:2016few} (black), DARWIN (purple)\cite{Aalbers:2016jon}. 
Thin solid black line shows the current lower bound on mass from direct searches at the LHC\cite{Aaboud:2017leg,CMS-PAS-SUS-16-048}. (b) Rescaled spin-dependent neutralino nucleon cross-section \sigsdp\ as a function of neutralino mass \mchi, for the a nearly pure higgsino in the CMSSM/mSUGRA. Solid lines show the 90\%~C.L. indirect upper bounds from IceCube\cite{Aartsen:2016zhm} (green) and Antares\cite{Adrian-Martinez:2016gti} (red). Dashed lines show projections for LZ\cite{Akerib:2015cja} (violet), XENON1T\cite{Aprile:2015uzo} (purple), Pico-500\cite{PICO} (blue), and DARWIN\cite{Aalbers:2016jon} (black).}
\label{fig:spinCS}
\end{figure}

We begin in \reffig{fig:spinCS}(a), where we plot the rescaled spin-independent neutralino-nucleon cross section versus neutralino mass for
a nearly pure higgsino under CMSSM/mSUGRA boundary conditions\cite{Kane:1993td}.\footnote{We remind the reader that this means scanning simultaneously over 4 free parameters: 
\mzero, the universal soft SUSY-breaking scalar mass at the GUT scale; \mhalf, the universal GUT-scale gaugino mass; \azero, 
the universal GUT-scale soft trilinear coupling; and \tanb, the ratio of the Higgs doublets' vevs. 
We scan them in this study over broad ranges: $\mzero,\mhalf\in[0.1\tev,30\tev]$, $\azero\in[-30\tev,30\tev]$, $\tanb\in[1,62]$. 
Additionally, one chooses the sign of $\mu$, which we set here to positive, as its sign does not much affect the region of parameter space with 
nearly pure higgsino DM (see, e.g.,\cite{Kowalska:2013hha,Roszkowski:2014wqa}). Note that the chosen input mass ranges encompass the parameter space region  
shown in \reffig{fig:spinCS} in its entirety. In it one finds $\mhalf\lesssim 0.6\,\mzero$, with $5\tev\lesssim \mzero\lesssim 25\tev$, $2.5\tev \lesssim \mhalf\lesssim 15\tev$ due to the Higgs mass measurement, see discussion on pages~14-15.}  
The color code depicts the higgsino DM relic abundance. For the points of the parameter space corresponding to \abund\ below the Planck measurement\cite{Ade:2015xua},
$\Omega_{\textrm{PL}}h^2 \approx 0.12$, we directly rescale \sigsip\ by $\xi=\abund/\Omega_{\textrm{PL}}h^2$,
assuming implicitly that the fraction of higgsino DM we measure locally today traces closely its early time large-scale 
freeze-out value. Solid tilted lines 
show recent direct upper bounds from the PandaX-II\cite{Tan:2016zwf} (maroon) and XENON1T\cite{Aprile:2017iyp} (blue) underground experiments. The latter is not much more constraining than an earlier bound from the now decommissioned LUX\cite{Akerib:2016vxi}. 
Dot-dashed lines show the projected reach of several upcoming and planned experiments.   

We also show in \reffig{fig:spinCS}(a) as a thin black line the current lower bound on mass from direct searches 
for compressed electroweakinos in final states with two low-momentum leptons at the LHC (Refs.\cite{Aaboud:2017leg,CMS-PAS-SUS-16-048}, 
following a proposal and case studies by\cite{Giudice:2010wb,Schwaller:2013baa}),
which is sensitive to higgsino DM for mass splitting $m_{\chi_2}-m_{\chi_1}=3-30\gev$.
One should also be aware of the estimated putative reach of the ILC in testing higgsinos\cite{Fujii:2017ekh}, 
which we do not show in the plot for lack of space. It extends to approximately 240\gev\ (480\gev), 
independently of mass splitting, if the beam energy is set to $s=(500\gev)^2$ ($s=1000^2\gev^2$).

In \reffig{fig:spinCS}(b) we show the rescaled spin-dependent neutralino-proton elastic scattering cross section, $\xi$\sigsdp, versus neutralino mass. 
We show with solid lines existing indirect upper bounds from observations of neutrinos from the Sun in the neutrino telescopes IceCube\cite{Aartsen:2016zhm} (green) and Antares\cite{Adrian-Martinez:2016gti} (red), interpreted for a predominantly $W^+W^-$ annihilation final state, which give a good approximation for the nearly pure higgsino case\cite{Roszkowski:2014iqa,Catalan:2015cna}. 
Dashed lines of different colors give various projections for the future direct reach in \sigsdp\
of underground detectors.

The relic density and DM observables are here calculated with \texttt{micrOMEGAs~v4.3.1}\cite{Belanger:2013oya}. 
The supersymmetric spectrum is calculated with \texttt{SPheno v4.0.3}\cite{Porod:2003um,Porod:2011nf}, and all model 
points are subject to LHC Higgs constraints from 
\texttt{HiggsSignals/HiggsBounds}\cite{Bechtle:2013xfa,Bechtle:2008jh,Bechtle:2011sb,Bechtle:2013wla} 
and to the Higgs mass measurement\cite{Aad:2015zhl}. 
The Higgs mass is calculated, like the SUSY spectrum, with the latest version of \texttt{SPheno}, 
which yields, in the regime where soft SUSY-breaking masses are well above $\sim 1\tev$,
a value in excellent agreement with other numerical packages, \texttt{SusyHD}\cite{Vega:2015fna} 
and \texttt{FlexibleSUSY}\cite{Athron:2016fuq}.
The calculated value is subject to an overall estimated theory uncertainty of approximately 2\gev\cite{Staub:2017jnp}, 
which we take into account in \reffig{fig:spinCS}.
Note that when the SUSY spectrum lies in the several~TeV regime or above, all electroweak precision and flavor observables,
including the anomalous magnetic moment of the muon, are expected to roughly maintain their SM value.

\begin{figure}[t]
\centering
\includegraphics[width=0.55\textwidth]{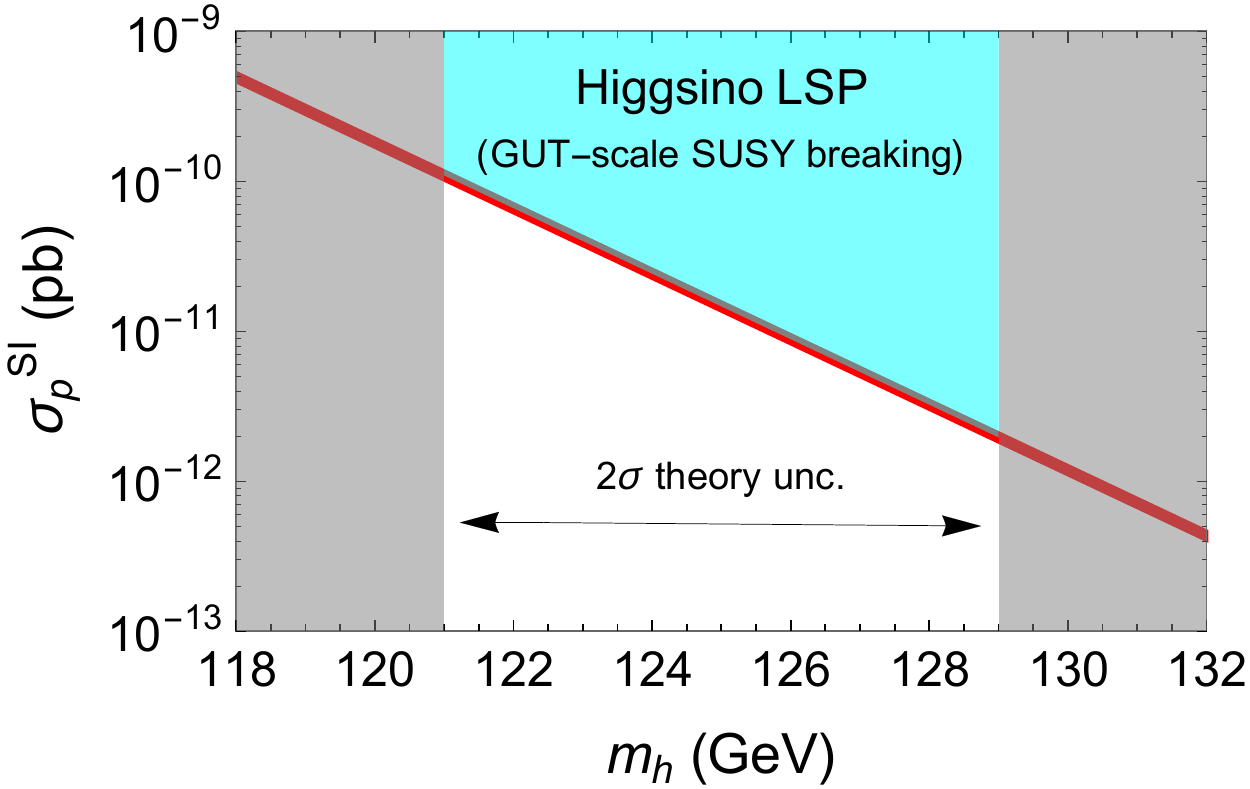}
\caption{Lower bound on \sigsip\ as a function of Higgs mass for a higgsino LSP of arbitrary mass
in generic models where the breaking of supersymmetry is transmitted at the GUT scale and the physical spectrum and 
EWSB are obtained after RGE to the low scale.}
\label{fig:higgscorr}
\end{figure}

We have chosen to show in \reffig{fig:spinCS} the higgsino parameter space under CMSSM boundary conditions, 
which provide a reasonable ansatz for models with scalar universality inspired by supergravity, and more 
generally cast in a lean framework scenarios in which supersymmetry breaking is transmitted 
to the visible sector at some high scale (the GUT scale) and EWSB is 
obtained radiatively around the minima of the MSSM scalar potential.
In models defined in this way 
one observes, for a higgsino-like neutralino, strong correlation 
between the Higgs boson mass and the allowed minimum value of \sigsip. We show this in \reffig{fig:higgscorr}, where we plot
the lower bound on \sigsip\ as a function of Higgs mass for a higgsino LSP of arbitrary mass. 
The correlation between minimum cross section and Higgs mass translates
in \reffig{fig:spinCS}(a) into a lower bound on \sigsip\ when $\mchi\approx1\tev$.

To qualitatively understand what is happening, let us recall from \refsec{sec:mixed} that in order to push down 
\sigsip\ for a predominantly higgsino-like neutralino 
one must increase purity $f_h$ or, in other words, raise the wino and bino masses, cf.~\refeq{hinofrac}. 
Very heavy winos/binos at the GUT scale feed through the RGE on the 
low-scale value of the soft SUSY-breaking up-type Higgs doublet mass, 
which carry SU(2) isospin and hypercharge, and also tend to push down the right-handed stop mass. 
This happens even in scenarios where the gluino mass 
is not universal and can be found relatively close 
to the higgsino, like those analyzed in\cite{Kowalska:2014hza}.
 
In order to keep the Higgs doublet soft mass under control, so to obtain 
a higgsino-like LSP after EWSB, and avoid tachyonic physical states, numerical scans are in this situation driven to
large negative \azero\ and/or larger soft scalar mass. 
Both solutions have the net effect of pushing up the Higgs boson mass 
and give rise to the behavior we observe in \reffig{fig:higgscorr}.\footnote{The attractiveness, from the phenomenological point of view,
of a lower bound on the neutralino scattering cross section determined by the Higgs mass measurement was pointed out early on 
in Bayesian analyses of the CMSSM/NUHM\cite{Kowalska:2013hha,Roszkowski:2014wqa,Roszkowski:2017nbc}.
The exact minimal cross section depends strongly
on the calculation of the Higgs mass itself, and on how it translates into mass predictions for the sparticles. 
In \texttt{SPheno~v4.0.3}, $\mhl\approx 125\gev$ leads to 
less optimistic expectations for the mean SUSY scale than in the  
versions of \texttt{SOFTSUSY}\cite{Allanach:2001kg} 
or \texttt{FeynHiggs}\cite{Hahn:2013ria} used in\cite{Kowalska:2013hha,Roszkowski:2014wqa}. 
Hence the parameter space in \reffig{fig:spinCS}(a) extends to 
lower \sigsip\ values than in those studies.}   

There is no apparent lower bound on the scattering cross section if we relax the requirement of radiative 
EWSB from boundary conditions generated at the GUT scale. 
This is the case, for example, in models where the typical mass of scalar particles is by several orders of magnitude 
decoupled from the electroweak vev (see, e.g.,\cite{Hall:2011jd,Fox:2014moa,Benakli:2015ioa}), and 
one does not expect to infer strict relations between the mechanism of SUSY-breaking and 
EWSB. The relic density alone determines then the mass of the higgsino-like DM, and purity $f_h$ can be extremely close to 1. 
We generically indicate with a black arrow in \reffig{fig:spinCS}(a) 
the parameter space for higgsino DM in those models, which can extend well below the neutrino background floor\cite{Hill:2013hoa,Nagata:2014wma}. 

This highly inaccessible part of the higgsino parameter space proves particularly tricky to probe. 
For underabundant higgsinos, $\mu\ll1\tev$, interesting venues for detections can be provided, for very small mass splitting, $m_{\chi^{\pm}}-\mchi\approx 150\mev$, by future collider searches for disappearing tracks\cite{Mahbubani:2017gjh,Fukuda:2017jmk}. 
If there is a sizable CP violating phase, future electron dipole moment experiments 
might be sensitive to parameter space with purity in excess of 99.99\%\cite{Nagata:2014wma}.
And possibly new venues for detection are given by the cooling curve of white dwarfs\cite{Krall:2017xij}. 
Additional opportunities for the future detection of higgsino-like compressed spectra, in particular for long-lived particles
with a relatively short lifetime, can arise then in electron-proton colliders\cite{Curtin:2017bxr}.
       
\begin{figure}[t]
\centering
\subfloat[]{
\includegraphics[width=0.46\textwidth]{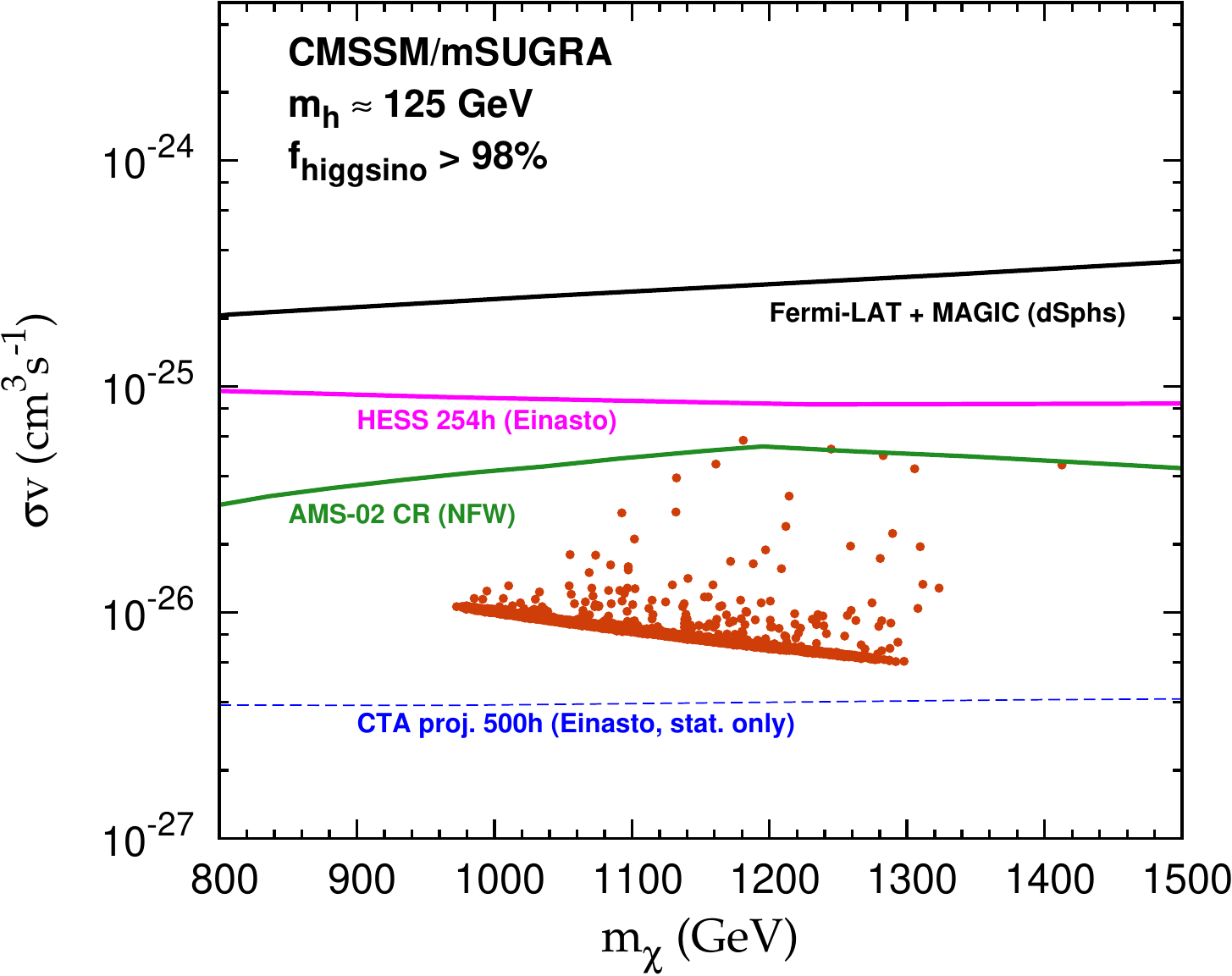}
}
\hspace{0.02\textwidth}
\subfloat[]{
\includegraphics[width=0.46\textwidth]{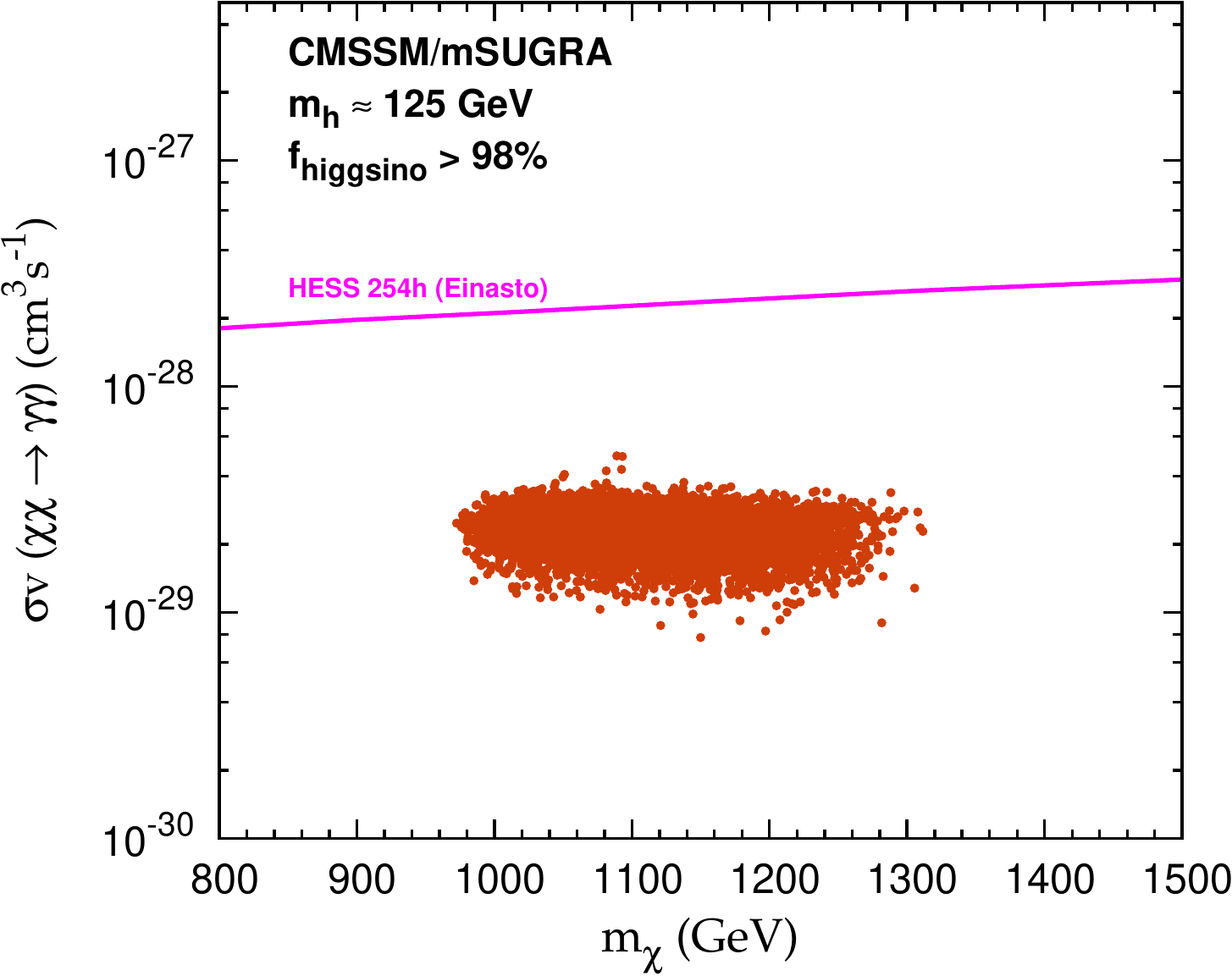}
}
\caption{(a) Indirect detection bounds and projections 
in gamma-ray searches in space and terrestrial telescopes for
$\sim 1\tev$ higgsino DM under CMSSM/mSUGRA boundary conditions.
Solid black line shows 90\%~C.L. upper bounds on the present-day annihilation cross section to $W^+W^-$ from the statistical 
combination of Fermi-LAT and MAGIC observations of dSphs\cite{Ahnen:2016qkx}; solid magenta line shows the recent
bound from 10-year observation of the Galactic Center at H.E.S.S.\cite{Abdallah:2016ygi} 
under the Einasto profile assumption; solid green line shows the upper bound from antiproton cosmic-ray (CR) data at AMS-02\cite{Aguilar:2016kjl} according to\cite{Cuoco:2017iax} 
for the NFW profile; 
and dashed blue line shows the projected reach of CTA 500h 
under the Einasto profile assumption\cite{Roszkowski:2014iqa}.
(b) In magenta, the current 95\%~C.L. upper bound on the annihilation cross section (times velocity) to gamma-ray lines, 
$\sigma_{\gamma\gamma} v$, from H.E.S.S.\cite{Rinchiuso:2017kfn} under the Einasto profile assumption, 
compared to the cross section of our $\sim 1\tev$ higgsino points.}
\label{fig:hinosigv}
\end{figure}

We finally show in \reffig{fig:hinosigv} the status of indirect detection bounds and projections 
in gamma-ray searches in space and terrestrial telescopes for
$\sim 1\tev$ higgsino DM under CMSSM/mSUGRA boundary conditions
(we implicitly assume that the chances for detection maximize if higgsinos saturate the relic abundance). 
In \reffig{fig:hinosigv}(a), solid black line shows the most recent 90\%~C.L. upper bound on the present-day \sigv\ from the statistical 
combination of Fermi-LAT and MAGIC observations of dSphs\cite{Ahnen:2016qkx}, and the magenta line draws the recent
bound from 10-year observation of the Galactic Center at H.E.S.S.\cite{Abdallah:2016ygi} 
under the Einasto profile assumption. We adopt the bounds in the $W^+W^-$ final state interpretation, which give a good 
approximation for the $\sim 1\tev$ higgsino. 

For the $W^+W^-$ final state we show in solid green the determination by\cite{Cuoco:2017iax} 
of the 95\%~C.L. upper bound on \sigv\ from antiproton CR data at AMS-02\cite{Aguilar:2016kjl}, under the NFW profile assumption. 
Note that the bound is subject to uncertainties related to the choice of diffusion model for CR propagation in the Galaxy. Some of these choices can in fact
weaken it\cite{Cuoco:2017iax}, and push it up to approximately the level of the H.E.S.S. limit. Finally,
dashed blue line shows the projected statistical reach of CTA 500h, 
under the Einasto profile assumption\cite{Roszkowski:2014iqa,Carr:2015hta}. 
Note that including the systematic uncertainty from diffuse astrophysical radiation will most likely weaken the extent of the projected 
reach\cite{Silverwood:2014yza,Catalan:2015cna}.
Also note in \reffig{fig:hinosigv}(a) that some model points are characterized by \sigv\ significantly above the thermal relic expectation, 
due to the presence of the heavy pseudoscalar Higgs mass at $\ma \approx 2\,\mchi$\cite{Roszkowski:2014wqa,Roszkowski:2014iqa}.
Regions of the parameter space that allow for this serendipitous coincidence thus see their indirect detection 
prospects improve significantly.  

We show in \reffig{fig:hinosigv}(b), as a magenta solid line, the current 95\%~C.L. upper bound on the annihilation cross section (times velocity) 
to gamma-ray lines from the final 254h data at H.E.S.S.\cite{Rinchiuso:2017kfn} under the Einasto profile assumption. The line is 
compared to the cross section of our $\sim 1\tev$ higgsino points, which lie well below the limit.

\subsection{The soft SUSY scale and fine tuning\label{sec:hinoHiggs}}

We conclude with a few words about the expected scale of the supersymmetric particles associated with higgsino DM. In truth, 
little is known in this regard, as the issue is highly model-dependent and there is not one only 
way of inferring the scale of SUSY breaking. 

Of course, expressions similar to Eqs.~(\ref{hinofrac})-(\ref{hinosigsip}) 
can give us a lower bound on the scale of the electroweak gauginos
for every given upcoming new constraint on \sigsip, 
but to be precise one should then take into account the rich parametric dependence of the full formulas.
Equivalently, the Higgs mass measurement tells us that in all likelihood stops and gluinos sit well above the LHC reach, 
but little more than that is known, as expectations depend strongly on parameters like \tanb\ and the trilinear coupling $A_t$.    

\begin{figure}[t]
\centering
\subfloat[]{
\label{fig:b}
\includegraphics[width=0.50\textwidth]{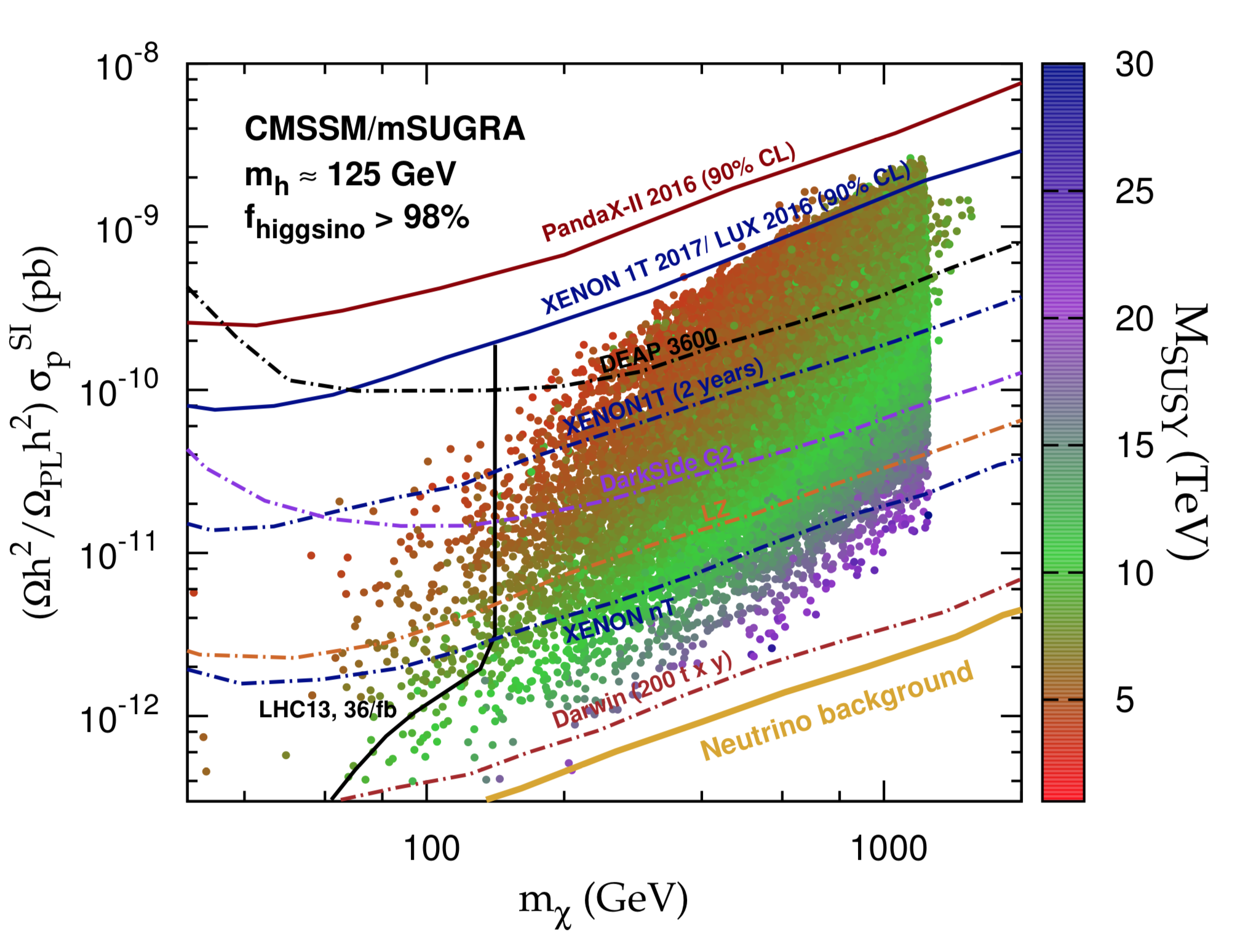}
}
\subfloat[]{
\label{fig:c}
\includegraphics[width=0.50\textwidth]{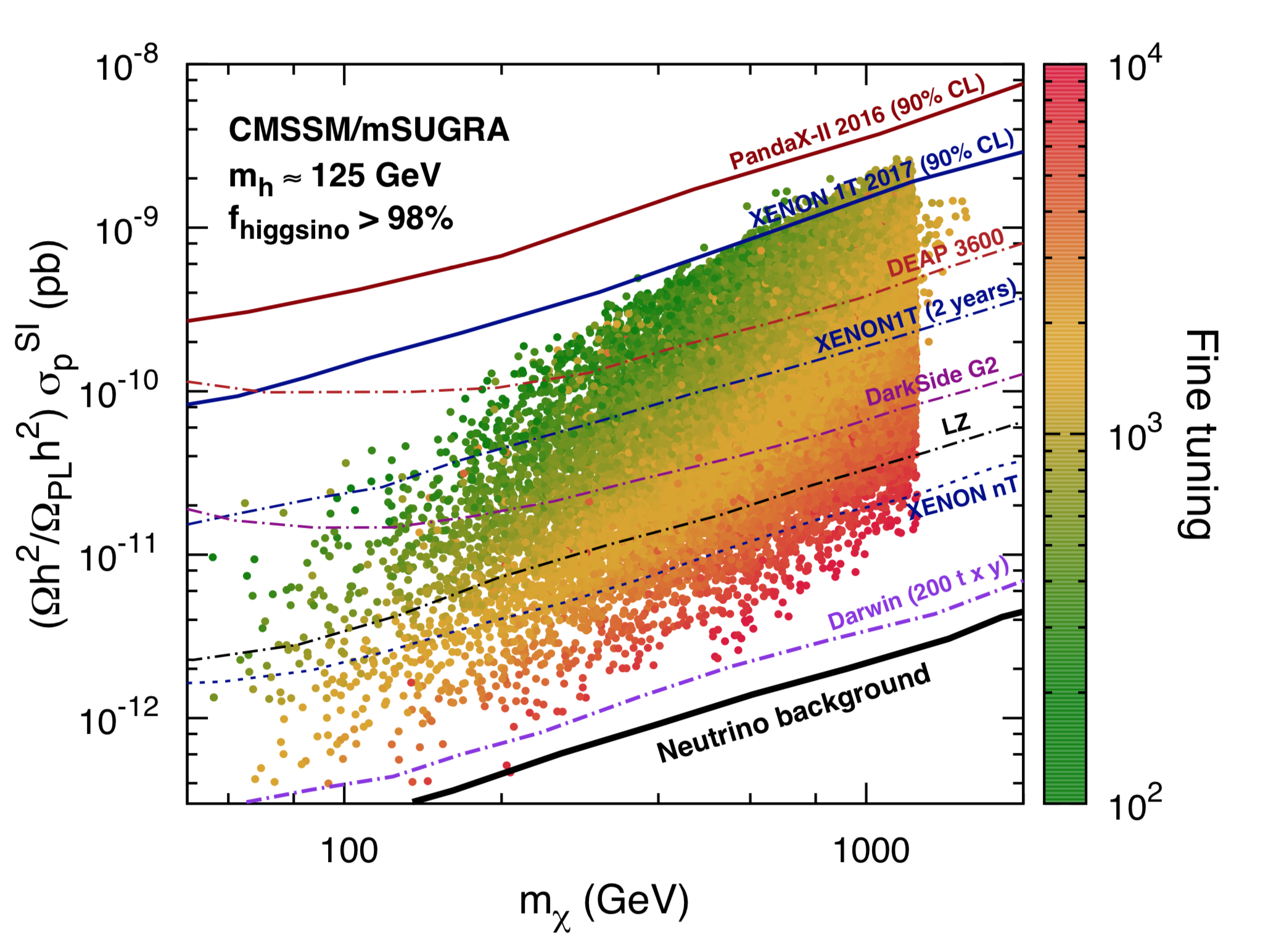}
}
\caption{(a) A plot of $\msusy=(\mstopone\,\mstoptwo)^{1/2}$ in the (\mchi, $\xi$\sigsip) plane with higgsino 
DM under CMSSM/mSUGRA boundary conditions. 
(b) EWSB fine tuning for points with higgsino DM in the (\mchi, $\xi$\sigsip) plane.}
\label{fig:msusy}
\end{figure}

Thus, without pretence of presenting any universally valid result, but to just show an example of a model where 
the measurement of the Higgs mass actually does provide predictions for the maximally allowed typical scale of the superpartners,
we present in \reffig{fig:msusy}(a) the distribution of the mean stop mass, $\msusy=(\mstopone\,\mstoptwo)^{1/2}$,  
under CMSSM/mSUGRA boundary conditions 
in the (\mchi, $\xi$\sigsip) plane with higgsino DM. One can see that by approximately 
the next round of XENON-1T data we will be starting to probe the 10\tev\ range of the superpartners if the DM
is entirely composed of higgsinos. Note also that, for higgsino mass $\mchi\lesssim 140\gev$, the LHC is already excluding,
with direct soft-lepton bounds on electroweakinos, the parameter space corresponding to $\msusy\lesssim 8-10\tev$. 

Finally, like all BSM models developed at least in part to deal with the hierarchy problem, 
after the first two runs of the LHC models with higgsino DM have become marred by a certain amount of EWSB fine tuning. 
The severity of this issue depends, of course, on the specific features of each model: how EWSB is obtained and 
the relation to the mass of the Higgs boson. In the context of the CMSSM, the fine tuning associated with higgsino DM 
is shown in \reffig{fig:msusy}(b), where we plot  in the (\mchi, $\xi$\sigsip) plane the size of the usual Barbieri-Giudice
measure\cite{Ellis:1986yg,Barbieri:1987fn} (following the prescription of\cite{Ross:2017kjc}).\footnote{We remind the reader that the Barbieri-Giudice measure is generally defined as $\max_{p_i} |\partial \log M_Z^2/\partial \log p_i|$, where the $p_i$ are the model's input parameters at the typical scale of the messengers for SUSY breaking. In the CMSSM these are the GUT-defined parameters \mzero, \mhalf, \azero, $B_0$, $\mu_0$.} 
No point shows EWSB fine tuning of less than a part in 100, as direct consequence of the Higgs mass measurement, 
and one can observe the well-known fact that higgsino points favored by expectations of naturalness 
correspond to $\mchi<1\tev$ and lead to $\abund\ll 0.12$.  
For the specific case of the $\sim 1\tev$ higgsino, a failure to observe a signal in, say, the next round of XENON-1T data will imply 
a fine tuning greater than one part in $10^3$, with rapid increase with each successive milestone exclusion.\footnote{There exist ways of embedding the MSSM in UV completions 
that can lead to lower fine tuning for higgsino DM, see, e.g.,\cite{Kowalska:2014hza,Ross:2016pml}.} 

However, we emphasize that a large fine tuning is by no means exclusive to the CMSSM, to higgsino DM, or even to SUSY in general
(see, e.g.,\cite{Barnard:2017kbb} for fine tuning in a non-SUSY scenario). 
As a matter of fact, the majority of phenomenological DM models found in the literature 
do not even attempt to construct a UV completion 
that could directly relate their free parameters to the physics of the high scale. 
It is very possible that once a discovery is finally made many of the suspended questions will start to find their answers. 
Higgsinos appear to be just in the perfect position to usher, in case of their eventual discovery, a new era of understanding.

\section{Summary and conclusions}\label{sec:sum}

The appealing theoretical features of the MSSM have made it, through the years, a natural favorite among 
the theoretical frameworks incorporating a possible DM particle. 
In this review, we have given a summary of the current status of phenomenological constraints on the DM candidates of 
the MSSM and have highlighted the growing consensus that, 
although available parameter space remains open for most DM aspirant particles, only one of them, 
the higgsino-like neutralino, is almost entirely free of tension from the increasing amount of observational data. 

Much of what makes higgsinos very attractive is the fact that the current constraints are not evaded 
with specific arrangements of some model parameters, 
but rather as a consequence only of the higgsino isospin quantum numbers, which lead to a fairly large mass to produce \abund\ in agreement with observations, and of the mass splittings among its neutral and charged components, which stem directly from EWSB.  
As these are not exotic features, however, 
one reasonably expects that the higgsino parameter space will not remain unexplored indefinitely. 

We have thus reviewed the excellent prospects for detection of higgsinos in the traditional experimental venues 
of direct DM detection in underground searches, indirect detection from astrophysical observations, and collider 
accelerators, all of which show reasons for optimism. 
The prospects are particularly enticing in supergravity-inspired scenarios with radiative EWSB,
where the overall consistency of the theoretical picture requires a lower bound on the spin-independent cross section for higgsinos, 
determined indirectly but convincingly by the measured value of the Higgs boson mass.

For those models that might instead be characterized 
by very large scales for the superpartners (in agreement with the 125\gev\ Higgs mass when \tanb\ is close to 1), 
the prospects for detection are more tricky to assess, but not without hope.  
We have drawn the reader's attention to a few references that promoted alternative venues for the explorations of this more 
fleeting scenarios. Promising venues are given by the experimental determination of dipole moments, 
disappearing track signatures in colliders, and the measurement of cooling curves in white dwarfs and neutron stars.

Overall, we hope this might serve as an agile but comprehensive report on the consistency of the higgsino DM picture, and on the 
multiple opportunities that arise for its observation in the not so distant future.
    
\bigskip
\begin{center}
\textbf{ACKNOWLEDGMENTS}
\end{center}
We would like to thank Luc Darm\'e for his comments on the manuscript and discussions.
The use of the CIS computer cluster at the National Centre for Nuclear Research in Warsaw is gratefully acknowledged. 
The authors declare that there is no conflict of interest regarding the publication of this article.
\bigskip
\bibliographystyle{utphysmcite}
\bibliography{KE3}

\end{document}